\documentclass[11pt,fleqn,leqno]{article}
\usepackage{graphicx}
\usepackage{a4}

\usepackage{amscd}
\usepackage{amsfonts}
\usepackage{amsmath}
\usepackage{amsthm}
\usepackage{amssymb}

\numberwithin{equation}{section}
\newtheorem{thm}{Theorem}[section]

\newtheorem{rem}[thm]{Remark}

\newtheorem{prop}[thm]{Proposition}
  {\noindent\textbf{Proof.}}%
  {\qed}

\def\const{\operatorname{const}}
\def\rank{\operatorname{rank}}
\def\pr{\partial}
\def\sfrac#1#2{{\textstyle\frac{#1}{#2}}}
\def\grad{\operatorname{grad}}
\def\rot{\operatorname{rot}}
\def\c{\operatorname{c}}
\def\Cal{\mathcal}
\def\R{{\mathbb R}}

\begin{document}

\title{The relativistic mechanics in a nonholonomic setting:\\
A unified approach to particles with non-zero mass 
\\ and massless particles.}

\author{Olga Krupkov\'{a}$^\dag$ and Jana Musilov\'{a}$^\ddag$}
\date{December 2008}
\maketitle

\begin{center}
$^\dag$ {\footnotesize\it Department of Algebra and Geometry, Faculty of Science, 
Palack\'y University\\
Tomkova 40, 779 00 Olomouc, Czech Republic
\\ 
e-mail:  krupkova@inf.upol.cz}
\\
$^\ddag$ {\footnotesize\it Faculty of Science, Masaryk University, Kotl\'{a}\v{r}sk\'{a} 2, 611 37 Brno,
Czech Republic\\
e-mail: janam@physics.muni.cz}

\end{center}

\abstract{A new approach to relativistic mechanics is proposed, suitable to describe dynamics of different kinds of relativistic particles. Mathematically it is based on an application of the recent geometric theory of nonholonomic systems on fibred manifolds. A setting based on a natural Lagrangian and a constraint on four-velocity of a particle is proposed, that allows a unified approach to particles with any (positive/negative/zero) square of mass. The corresponding equations of motion are obtained and discussed. In particular, new forces are found (different from the usual Lorentz force type term), arising due to the nonholonomic constraint. A possible meaning and relation with forces previously proposed by Dicke is discussed. In particular, equations of motion of tachyons and of massless particles are studied and the corresponding dynamics are investigated.
 \\

\\

\noindent{\sc MS classification.}
70G45, 70G75, 37J60, 70F25


\section{Introduction}

In this paper we are concerned with a mathematical model leading to a generalised setting of relativistic dynamics.   

As it is well-known, within the framework of the special relativity theory, a point particle of a constant rest mass $m_0 > 0$ moving in an electromagnetic field $(\vec{A}, V)$ on $\R^4$ is usually described by the 
action integral $\int_a^b \Cal L {\rm d}t$ with the Lagrangian
\begin{equation} \label{standL}
\Cal L =-m_0 c^2 \sqrt{1-\frac{v^2}{c^2}}+ \frac{e}{c} \vec A\vec v - eV,
\end{equation}
providing motion equations
\begin{equation}
\frac{\rm d}{{\rm d}t}\Bigl(\frac{m_0 \vec{v}}{\sqrt{1-\frac{v^2}{c^2}}}\Bigr)
=  \frac{e}{c} (\vec{v}\times{\rm rot}\vec{A}) - \frac{e}{c} \frac{\partial\vec{A}}{\partial t}
- e \,{\rm grad}V,
\end{equation}
with the Lorentz force, $\vec{\Cal F}_L$, on the right-hand side.
More generally, it is known (and rather surprising) that equations
\begin{equation}
\frac{\rm d}{{\rm d}t}\Bigl(\frac{m_0 \vec v}{\sqrt{1 - \frac{v^2}{c^2}}}\Bigr) = \vec{\Cal F}
\end{equation}
are variational ``as they stand" if and only if the force on the right-hand side is a {\em Lorentz-type force},
i.e., $\vec{\Cal F}$ = $\vec{\Cal F}_L$,
for some functions $\vec{A}, V$ on $\R^4$ (\cite{Nov}, see also \cite{Kr-a97}).\footnote{Of course, in this general setting, the meaning of functions $\vec{A}, V$ need not be just a vector and scalar potential of an electromagnetic field. It includes also e.g. forces of inertia.}
Mathematically this result means that Lorentz-type force interactions are the only admissible interactions for a massive special-relativistic particle, {\em compatible with a variational principle for curves in $\R^3$ parametrised by time $t \in \R$}.
On the other hand, there appeared hypotheses that also different kinds of interactions should be possible: a significant prediction of such a non-Lorentz-type force is due to Dicke \cite{dicke}, however, a formula as well as exact arguments are missing up to now. Hence, a question arises on a proper general mathematical setting, providing equations of motion for particles in the Minkowski space-time, and clarifying admissible forces. 

A problem of this kind can be tackled by tools of the modern calculus of variations.
The above variational principle is apparently too restrictive: one should better consider a variational principle for one-dimensional {\em submanifolds in $\R^4$}. However, there is another interesting possibility, within an even more general framework, proposed and studied in \cite{krupkova-musilova-rel}: One can utilize a variational principle for {\em curves in the Minkowski space-time $(\R^4,g)$} together with the relativistic {\em constraint on the four-velocity}: from the mathematical point of view this is a {\em nonholonomic constraint}. The constraint defines a genuine {\em evolution space}, $\Cal Q$, which is a {\em submanifold in $\R \times \R^4 \times \R^4$} of codimension one. Then dynamics proceed in $\Cal Q$ and are governed by so-called {\em reduced Chetaev equations} \cite{sarlet, krupkova-JMP} that become the desired equations of motion for the problem in question. In the above mentioned paper
\cite{krupkova-musilova-rel} we applied the procedure directly to ``conventional" massive particles, moving in an electromagnetic and scalar field, and obtained a generalised formula for the force $\vec{\Cal F}$ covering also ``Dicke-type" interactions.
 
In this paper we show that ideas of this kind can be applied to study all kinds of point particles
admitted by the special relativity theory. We propose a unified mathematical setting suitable not only for ``conventional" particles (with a positive square of mass), but also for tachyons (particles with a negative square of mass), and particles with zero mass, moving in an electromagnetic and scalar field. The idea of  \cite{krupkova-musilova-rel} to tackle this problem as a variational problem with a nonholonomic constraint remains, however, the Lagrangian
and the constraint are appropriately modified: the Lagrange function we propose is {\em universal} for all
the kinds of particles (not containing the particle's mass), however, {\em particles are distinguished by the constraint} (mass appears in the constraint condition). In the next two sections of the present paper we explain our approach in general, in the rest we study in detail each kind of the particles separately. 
As the main result we obtain the corresponding equations of motion, and formulas for admissible forces. We discuss the meaning of the constraint condition, and of the new forces  (different from the usual Lorentz force type term), arising due to the nonholonomic constraint, and find among them a force complying with Dicke's predictions. In particular, we pay attention to equations of motion of tachyons and of massless particles, and investigate the corresponding dynamics. In this context one should mention at least two striking properties of massless particles: First, it turns out that dynamics of these particles are singular, not obeying the Newton's determinism principle, i.e. these particles
belong to mechanical systems with internal constraints studied (in the Lagrangian case) by Dirac.
Second, we found out that, if massless particles are admitted to move in a scalar field, this field does not influence on their dynamics. 
The last section is devoted to a summary of results of this paper, where also other interesting properties
of massive and massless particles, obtained within our approach, are recalled and discussed.


\section{Mathematical background}

From the point of view of mathematics, the main idea is to join variational approach, represented by the calculus of variations on fibred manifolds
(see \cite{Kr73, Kr08} or \cite{Kr-a97}) with nonholonomic geometric mechanics in jet bundles due to \cite{krupkova-JMP}, \cite{krupkova-RMP} (for alternative approaches see e.g. \cite{CaRa, leon-et-al, Gi, KoMa, massa-pagani, sarlet-et-al-1, sarlet-et-al-2} and others). 

Throughout the paper we shall use Einstein summation convention. Summation over greek (respectively, latin) indices proceeds from $1$ to $4$ (respectively, $1$ to $3$). 

First, let us recall a necessary mathematical background, adapted to our situation. Consider the manifold $\R^4$ with canonical coordinates $(q^\sigma)$, $1 \leq \sigma \leq 4$, endowed with the Minkowski metric field 
\begin{equation}
g = \delta_{kl} \, {\rm d}q^k \otimes {\rm d}q^l - {\rm d}q^4 \otimes {\rm d}q^4.
\end{equation}
Curves in $\R^4$ will be represented by their graphs, i.e. {\em sections of the fibred manifold $\pi: \R \times \R^4 \to \R$}. If $\c: s \to \c(s)$ is a curve in $\R^4$ parametrised by a real parameter $s \in \R$, then the corresponding section is the mapping $\gamma: \R \ni s \to \gamma(s) = (s, \c(s)) = (s, q^\sigma(s)) \in \R \times \R^4$. We shall need the manifold
\begin{equation}
J^1 \pi = \R \times T\R^4 = \R \times \R^4 \times \R^4
\end{equation}
(the first jet prolongation of $\pi$), called {\em evolution space}. We shall consider it with {\em fibred coordinates} $(s, q^\sigma, \dot q^\sigma)$; note that the ``dot" coordinates are defined by 
\begin{equation}
\dot q^\sigma \circ J^1\gamma = \frac{\rm d}{{\rm d} s} (q^\sigma \circ \gamma),
\end{equation}
for every section $\gamma$ of $\pi$. 

On $J^1\pi$ we have the so-called {\em contact structure}, locally generated by basic contact $1$-forms
\begin{equation}
\omega^\sigma = {\rm d}q^\sigma - \dot q^\sigma {\rm d}s, \quad 1 \leq \sigma \leq 4.
\end{equation}

In what follows, we shall often consider vector fields and $1$-forms, defined on the Minkowski space-time $\R^4$ (i.e., on the typical fibre of the fibration $\pi$), and call them {\em contravariant} and {\em covariant four-vector fields}, respectively. In fibred coordinates a contravariant, respectively, covariant vector field reads
\begin{equation}
\hat u = \hat u^\nu \frac{\pr}{\pr q^\nu}, \quad \text{respectively,} \quad
\phi = \phi_\nu {\rm d}q^\nu,
\end{equation}
where the components $\hat u^\nu$ and $\phi_\nu$, $1 \leq \nu \leq 4$, are functions on $\R^4$, i.e. depend upon the coordinates $(q^\sigma)$. Due to the structure of the evolution space $J^1\pi$ (being a product of $\R$ and the tangent space $T\R^4 = \R^4 \times \R^4$), contravariant four-vector fields are sections of the bundle $T\R^4 \to \R^4$, i.e., it holds $q^\nu \circ \hat u = q^\nu$ and
\begin{equation}
\dot q^\nu \circ \hat u = \hat u^\nu (q^1, q^2, q^3, q^4), \quad 1 \leq \nu \leq 4.
\end{equation} 
Covariant four-vector fields are sections of the cotangent bundle $T^*\R^4 \to \R^4$.

A {\em Lagrangian} is defined to be a differential one-form on the evolution space $J^1 \pi$, horizontal with respect to the projection onto the base $\R$. 
In fibred coordinates, a Lagrangian is expressed by 
$\Lambda = L(s, q^\sigma, \dot q^\sigma) \, {\rm d}s$, and {\em action} (over an interval 
$[a,b] \subset \R$) is a function
\begin{equation}
S: \Gamma(\pi) \ni \gamma \to \int_{[a,b]} J^1\gamma^* \Lambda = 
\int_a^b (L \circ J^1\gamma) \, {\rm d}s  \in \R,
\end{equation}
where $\Gamma(\pi)$ denotes the set of sections of $\pi$ with domains of definition containing the interval $[a,b]$, and $J^1\gamma$ is the first jet prolongation of $\gamma$, i.e. a section of $J^1\pi$ defined by $J^1\gamma(s) = (s, \c(s), {\rm dc}/{\rm d}s) = (s, q^\sigma(s), \dot q^\sigma(s))$. 

It is important to notice the geometric meaning of the Euler--Lagrange equations of a Lagrangian defined on $J^1\pi$. Extremals of $\Lambda$ are {\em integral sections of a distribution on the evolution space} $J^1\pi$ (by a distribution we mean a subbundle of the tangent bundle to the manifold $J^1\pi$). This distribution, called {\em dynamical distribution}, arises as the {\em characteristic distribution of the $2$-form}
\begin{equation}
\alpha=A_\sigma\omega^\sigma\wedge\,{\rm d}s+B_{\sigma\nu}\omega^\sigma\wedge\,{\rm
d}\dot{q}^\nu,
\end{equation}
where
\begin{equation}
A_\sigma = \frac{\pr L}{\pr q^\sigma} - \frac{\pr^2 L}{\pr s \, \pr \dot q^\sigma} - \frac{\pr^2 L}{\pr q^\nu \pr \dot q^\sigma} \dot q^\nu,
\quad
B_{\sigma\nu} = - \frac{\pr^2 L}{\pr \dot q^\sigma \pr \dot q^\nu}.
\end{equation}
In this notation, Euler--Lagrange equations of $\Lambda$ read 
\begin{equation}
A_\sigma + B_{\sigma\nu} \ddot q^\nu = 0, \quad 1 \leq \sigma \leq 4,
\end{equation}
or, in intrinsic form, $J^1 \gamma^* i_\xi\alpha = 0$ for every vector field $\xi$ on $J^1\pi$.
The rank of the dynamical distribution may be greater than one. It is equal to one if and only if the matrix 
$(B_{\sigma \nu})$ is {\em regular}, i.e. if
the Lagrangian is regular; in this case trajectories in the evolution space are integral curves of a {\em single vector field}. For more details on the structure of solutions of Euler--Lagrange equations and corresponding integration methods for both  regular and non-regular Lagrangians we refer to 
\cite{Kr-a97} and \cite{KruPri}.

In this paper we shall consider Lagrangians where the Lagrange function $L$ does not depend explicitly upon the parameter $s$, as relevant from the physical point of view. In this case we have
\begin{equation}
(L\circ J^1 \gamma) (s) = L \Bigl(\c(s), \frac{\rm dc}{{\rm d}s}\Bigr) = L (q^\sigma(s), \dot q^\sigma(s)) = 
(L \circ \hat u \circ \c)(s) = (L \circ \hat u)(\c(s)),
\end{equation}
hence, {\em we can consider a Lagrange function defined on $\R \times T \R^4$, of the form
$L = L(\hat u)$}.

A {\em nonholonomic constraint} is a submanifold $\Cal Q$ of $J^1\pi$, fibred over $\R \times \R^4$. Whenever convenient, we denote by $\iota$ the canonical embedding of $\Cal Q$ into $J^1\pi$.
We shall consider a nonholonomic constraint of codimension $1$. Such a constraint is given by one
first-order partial differential equation
\begin{equation}
f(s,q^\sigma, \dot q^\sigma) = 0,
\end{equation}
such that 
\begin{equation} \label{rank}
\rank (\pr f/\pr \dot q^\sigma) = 1.
\end{equation}
By the rank condition, we may assume that the constraint is defined by equation ``in normal form"
\begin{equation}
\dot q^4 = h(s, q^\sigma, \dot q^l),
\end{equation}
where $1 \leq \sigma \leq 4$, and $1 \leq l \leq 3$ (i.e. that $f \equiv \dot q^4 - h$).  The given fibred coordinates induce on the submanifold $\Cal Q$ adapted coordinates 
$(s, q^1, q^2, q^3, q^4, \dot q^1, \dot q^2,\dot q^3)$. 

A constraint in the evolution space $J^1\pi$ gives rise to a {\em constraint structure} on $J^1\pi$, generated  by the $1$-form
\begin{equation}
\varphi = f {\rm d}s + \frac{\pr f}{\pr \dot q^\sigma} \omega^\sigma.
\end{equation}
As shown in \cite{krupkova-JMP, massa-pagani}, due to the constraint structure, the constraint  $\Cal Q$ is naturally endowed with a distribution (subbundle of the tangent bundle to $\Cal Q$)  called {\em canonical distribution}, annihilated by the one-form
\begin{equation}
\bar \varphi \equiv \iota^* \varphi =  -\frac{\partial h}{\partial\dot{q}^l}\omega^l + ({\rm d}q^4 - h\, {\rm d}s)
\end{equation}
on the manifold $\Cal Q$. The canonical distribution represents admissible directions in the evolution space (a nonholonomic  ``principle of virtual displacements").

There are two models for describing a constrained system, both will be useful for our further considerations. The first one is more traditional, describing the constrained system as a {\em deformation} of the original unconstrained system due to a {\em constraint force}, naturally generated by the constraint. The deformed system is thus a {\em  new mechanical system defined on $J^1\pi$}. It is represented by the $2$-form $\alpha_C = \alpha - \Phi$,
where $\alpha$ represents the original mechanical system and 
\begin{equation}
\Phi = \lambda \, \varphi \land {\rm d}s =  \lambda \, \frac{\pr f}{\pr \dot q^\sigma} \omega^\sigma
\land {\rm d}s
\end{equation}
is the constraint force, called {\it Chetaev force}.
Constrained trajectories are then integral curves of the characteristic distribution of the $2$-form 
$\alpha_C$, passing in the manifold $\Cal Q$. Equations for these curves, called {\em Chetaev equations} \cite{Chet}, depend upon one Lagrange multiplier $\lambda$ (to be determined), and read
\begin{equation}
\frac{\pr L}{\pr q^\sigma} - \frac{\rm d}{{\rm d}s} \frac{\pr L}{\pr \dot q^\sigma} = 
\lambda \, \frac{\pr f}{\pr \dot q^\sigma}, \quad
1 \leq \sigma \leq 4. 
\end{equation}
The second model is more geometrical. It represents a non-holonomic mechanical system as an object {\em defined on the constraint} $\Cal Q$. Hence, {\em the manifold $\Cal Q$ has the meaning of the evolution space for the constrained system}. Concretely, the constrained system is represented by the class of $2$-forms $[\bar \alpha] = \iota^*\alpha + [\bar \varphi]$ {\em on $\Cal Q$}, where $[\bar \varphi]$ is the ideal generated by the constraint form $\bar \varphi$. Now, admissible trajectories are integral sections of the canonical distribution, and equations of motion, called {\em Chetaev reduced equations} are equations for admissible trajectories that are integral curves of the characteristic distribution of 
$\bar \alpha$. Keeping notations used so far, they take the form
\begin{equation}
\label{reduced-equations}
 \bar{A}_l+\bar{B}_{ls}\ddot{q}^{s} = 0, \quad 1 \leq l \leq 3,
\end{equation}
where
\begin{equation} \label{red-A}
\bar{A}_l=\left(A_l+ A_4 \frac{\partial h} {\partial\dot{q}^l} + \left(B_{l4}+
B_{44}\frac{\partial h}{\partial\dot{q}^l}\right)\right. \left.\left(\frac{\partial h}{\partial t}+
\frac{\partial h}{\partial
q^\sigma}\dot{q}^\sigma\right)\right)\circ\iota,
\end{equation}
\begin{equation}
\label{red-B}
\bar{B}_{ls}=\left(B_{ls} +2 B_{l4}
\frac{\partial h}{\partial\dot{q}^s} +\right.
\left.B_{44} \frac{\partial h}{\partial\dot{q}^l}\frac{\partial h} {\partial \dot{q}^s}\right)\circ\iota\,.
\end{equation}
It is important to stress that if the constraint is non-integrable (and this will be our case) the non-holonomic equations of motion {\em do not come from the ``constrained Lagrangian"} $\bar L = L \circ \iota = \bar L(s, q^\sigma, \dot q^l, h)$.


\section{A geometric setting for SRT particle dynamics}

In this paper we propose a model for particle dynamics on the Minkowski space, suitable for a unified treatment of all kinds of point particles (``classical" - with positive square of mass, tachyons - with negative square of mass, and particles with zero rest mass).

This model treats a particle as a Lagrangian system on a fibred manifold $\pi: \R \times \R^4 \to \R$ (where $\R^4$ is endowed with the Minkowski metric), with a non-holonomic constraint in the evolution space $J^1\pi = \R \times T\R^4$, and is based on the following two axioms:

\smallskip

\textbf{ (1) The Lagrange function is universal for all particles, and is polynomial in the ``four-velocity"}
\begin{equation}
\label{lagrangian-2}
\begin{aligned}
L(\hat u) &= - \sfrac{1}{2} g(\hat u,\hat u) + \phi(\hat u) - \psi 
= - \sfrac{1}{2} g_{\sigma \nu} \hat u^\sigma \hat u^\nu + \phi_\sigma \hat u^\sigma - \psi
 \\
&=  - \sfrac{1}{2}\Bigl((\dot{q}^4)^2-\sum_{p=1}^3(\dot{q}^p)^2
\Bigr) +\dot{q}^\sigma\phi_\sigma-\psi,
\end{aligned}
\end{equation}
where $\phi_\sigma$, $1 \leq \sigma \leq 4$, and $\psi$ are functions on the space-time $\R^4$.

Here ``universality" means that the Lagrangian {\em does not contain the particle mass}. 
Note that coefficients in the Lagrangian have the physical meaning of {\em admissibe fields}: behind the {\em $(2,0)$-tensor field} $g$, the Lagrangian contains a {\em covariant four-vector field} $\phi$ and a {\em scalar field} $\psi$.

\smallskip

\textbf{(2) A particle is determined by a non-holonomic constraint condition on the ``four-velocity"}: 
\begin{equation}
\label{relconstraint}
g(\hat u, \hat u) = \Cal M c^2,
\end{equation}
where $\Cal M = \Cal M (q^1,q^2,q^3,q^4)$ is a function on the space-time $\R^4$. 

The value of the function $\Cal M$ may vary from point to point; at a point in $\R^4$ it may be {\em positive, negative} or {\em  zero}.

Note that if we assume $\Cal M$ be {\em continuous} then the following property holds:  If 
$\Cal M(x) \ne 0$ at a point $x \in \R^4$ then there is an open neighbourhood $U$ around $x$ such 
that $\Cal M \ne 0$ on $U$. This means that if $\Cal M$ is positive (negative) at a point, it is positive
(negative) in a certain {\em neighbourhood} of that point.

\smallskip

Since $g(\hat u, \hat u) = ((\dot{q}^4)^2-\sum_{p=1}^3(\dot{q}^p)^2) \circ \hat u$, condition (\ref{relconstraint}) defines a {\em nonholonomic constraint}
\begin{equation}
\Cal Q: \quad (\dot{q}^4)^2-\sum_{p=1}^3(\dot{q}^p)^2 = \Cal M c^2
\end{equation}
in $J^1\pi = \R \times T\R^4$ if the rank condition (\ref{rank}) is satisfied, i.e., if
\begin{equation}
\rank (-\dot q^1, - \dot q^2, - \dot q^3, \dot q^4) = 1.
\end{equation}
Hence we have to exclude from $\R \times T\R^4$ the points where $\hat u =0$ (the zero section in 
$T\R^4$). However, we need more, namely, the constraint $\Cal Q$ be expressible in normal form
\begin{equation}
\dot q^4 = h(s,q^\sigma, \dot q^1,\dot q^2, \dot q^3).
\end{equation}
This is the case if 
\begin{equation} \label{rankcond}
\dot q^4 \ne 0.
\end{equation}
Excluding the hyperplane $\Cal H_0$ of points in $J^1\pi$ where $\dot q^4 = 0$ we get a
{\em disconnected open submanifold of the evolution space $J^1\pi$, consisting from two connected components}, $J^1\pi_{+}$ and $J^1\pi_{-}$, of points where $\dot q^4  > 0$  and $\dot q^4  < 0$, respectively.
On the components we have the induced global fibred coordinates $(s,q^\sigma, \dot q^\sigma)$,
and the constraint condition (\ref{relconstraint}) determines a constraint submanifold 
$\Cal Q \subset J^1\pi \backslash \Cal H_0 = \Cal Q_{+} \cup \Cal Q_{-}$, defined by the following equations in normal form:
\begin{equation} \label{Q+}
\Cal Q_{+} \subset J^1\pi_{+}:  \quad    \dot q^4 = \sqrt{\Cal M c^2 + \sum_{p=1}^3(\dot{q}^p)^2},
\end{equation}
\begin{equation}
\Cal Q_{-} \subset J^1\pi_{-}:   \quad    \dot q^4 = - \sqrt{\Cal M c^2 + \sum_{p=1}^3(\dot{q}^p)^2}.
\end{equation}

In what follows {\em we shall choose for the evolution space of the constrained system the manifold
$\Cal Q_{+}$}.

Lagrangian (\ref{lagrangian-2}) gives rise to four Euler--Lagrange equations
\begin{equation}
\frac{\pr L}{\pr q^\sigma} - \frac{\rm d}{{\rm d}s} \frac{\pr L}{\pr \dot q^\sigma} = 0
\end{equation}
for graphs of curves $\c(s) = (q^\sigma(s))$ in the space-time $\R^4$, having the following form:
\begin{equation} \label{origeq}
\begin{aligned}
- \ddot{q}^l &+\dot{q}^\sigma\left(\frac{\partial\phi_
\sigma}{\partial q^l}-\frac{\partial\phi_l}{\partial
q^\sigma}\right) - \frac{\partial\psi}{\partial q^l} = 0, 
\\
\ddot{q}^4 &+\dot{q}^\sigma\left(\frac{\partial\phi_\sigma}
{\partial q^4}-\frac{\partial\phi_4}{\partial q^\sigma}\right) -
\frac{\partial \psi}{\partial q^4} = 0.
\end{aligned}
\end{equation}

{\em Equations we are looking for concern motions proceeding in the evolution space $\Cal Q_{+}$}.
Substituting
\begin{equation}
\begin{aligned}
A_l &= \dot{q}^\sigma\left(\frac{\partial\phi_
\sigma}{\partial q^l}-\frac{\partial\phi_l}{\partial
q^\sigma}\right) -\frac{\partial \psi}{\partial q^l},
\quad 
B_{lj} = - \delta_{lj}, 
\quad 1\leq l,j \leq 3,
\\
A_4 &= \dot{q}^\sigma\left(\frac{\partial\phi_\sigma}{\partial q^4}-\frac{\partial\phi_4}{\partial
q^\sigma}\right) -\frac{\partial \psi}{\partial q^4},
\quad
B_{4l}=B_{l4}=0, \quad B_{44}= 1,
\end{aligned}
\end{equation}
and
\begin{equation}
h = \sqrt{\Cal M c^2 + \sum_{p=1}^3(\dot{q}^p)^2}.
\end{equation}
into (\ref{red-A}) and (\ref{red-B}) we obtain
\begin{equation}
\begin{aligned}
\bar{A}_l &= \dot{q}^i\left(\frac{\partial\phi_i}{\partial q^l}-
\frac{\partial\phi_l}{\partial q^i}\right) -
\frac{\partial\psi}{\partial q^l}+
\left(\dot{q}^i\left(\frac{\partial\phi_i}{\partial q^4} -
\frac{\partial\phi_4}{\partial q^i}\right) -
\frac{\partial\psi}{\partial q^4}\right)
\frac{\dot{q}^l}{\sqrt{\Cal M c^2 + \sum_{p=1}^3(\dot{q}^p)^2}} 
\\
&+ \sqrt{\Cal M c^2 + \sum_{p=1}^3(\dot{q}^p)^2} \, \left(\frac{\partial\phi_4}{\partial
q^l}-\frac{\partial\phi_l}{\partial q^4} \right).
\\
\bar{B}_{lj} &= - \delta_{lj}+\frac{\dot{q}^l\dot{q}^j}
{\Cal M c^2 + \sum_{p=1}^3(\dot{q}^p)^2}.
\end{aligned}
\end{equation}
The desired equations are {\em three} equations for sections of the fibred manifold
$\Cal Q_{+}$ over $\R$, i.e., for curves $\c(s) = (q^\sigma(s))$ passing  in the space-time $\R^4$ and satisfying the constraint condition (\ref{Q+}), as follows:
\begin{equation} \label{geneq}
\bar{B}_{lj}\ddot{q}^j = - \bar{A}_l, \quad 1 \leq l \leq 3.
\end{equation}

Now, we have to distinguish two cases:

\subsection{$\Cal M = 0$ at a point $x \in \Cal Q_{+}$} 

If $\Cal M (x) = 0$ then the matrix $(\bar{B}_{lj})$ is {\em singular} at $x$, hence {\em motion equations (\ref{geneq}) cannot be put into the normal form}; explicitly they read
\begin{equation} 
\begin{aligned}
\Bigr(\delta_{lj} - \frac{\dot{q}^l\dot{q}^j} {\sum_{p=1}^3(\dot{q}^p)^2} \Bigl) \ddot{q}^j 
&= \dot{q}^i\left(\frac{\partial\phi_i}{\partial q^l} -
\frac{\partial\phi_l}{\partial q^i}\right) +
 \frac{\dot{q}^l\dot{q}^i}{\sqrt{\sum_{p=1}^3(\dot{q}^p)^2}} 
\left(\frac{\partial\phi_i}{\partial q^4} -
\frac{\partial\phi_4}{\partial q^i}\right)
\\
& + \sqrt{\sum_{p=1}^3(\dot{q}^p)^2} \, \left(\frac{\partial\phi_4}{\partial
q^l}-\frac{\partial\phi_l}{\partial q^4} \right) -
\frac{\partial\psi}{\partial q^l} - \frac{\partial\psi}{\partial q^4}
\frac{\dot{q}^l}{\sqrt{\sum_{p=1}^3(\dot{q}^p)^2}},
\end{aligned}
\end{equation}
where $1 \leq l \leq 3$. Extremals of the constrained problem are curves 
$\gamma(s) = (s, q^\sigma(s))$ satisfying the above motion equations and the equation of the constraint $\Cal Q_{+}$
\begin{equation} 
\dot q^4 = \sqrt{\sum_{p=1}^3(\dot{q}^p)^2}.
\end{equation}
The dynamics proceed in the evolution space $\Cal Q_{+}$.

We shall deal with these equations in more detail later when we shall study massless particles.

\subsection{$\Cal M \ne 0$ at a point $x \in \Cal Q_{+}$} 

If $\Cal M (x) \ne 0$ then the matrix $(\bar{B}_{lj})$ is {\em regular} at $x$, and, due to continuity, it is regular in a neighbourhood $U$ around $x$. Hence, on $U$, {\em equations (\ref{geneq}) can be put into the 
normal form}
\begin{equation}
\label{gennormal}
\ddot{q}^j = {\cal{F}}^j.
\end{equation}
The inverse matrix $\bar{B}^{-1}=(\bar{B}^{jl})$ takes the form
\begin{equation}
\bar{B}^{-1}= - \frac{1}{\Cal M c^2}\left(
\begin{array}{ccc}
\Cal M c^2 + (\dot{q}^1)^2 & \dot{q}^1\dot{q}^2 & \dot{q}^1\dot{q}^3 \\
& & \\  \dot{q}^1\dot{q}^2 & \Cal M c^2 + (\dot{q}^2)^2 &
\dot{q}^2\dot{q}^3 \\ & & \\
\dot{q}^1\dot{q}^3 & \dot{q}^2\dot{q}^3 & \Cal M c^2 + (\dot{q}^3)^2 \\
\end{array}
\right),
\end{equation}
hence we obtain
\begin{equation}
\begin{aligned} \label{force4}
{\cal{F}}^j &= - {\cal{A}}^j = - \bar{B}^{jl}\,\bar{A}_l =
\frac{1}{\Cal M c^2}(\Cal M c^2 \delta^{jl}+\dot{q}^j\dot{q}^l)\,\bar{A}_l
\\
&=
\delta^{jl}\dot{q}^i\left(\frac
{\partial\phi_i}{\partial q^l}-\frac{\partial\phi_l}{\partial
q^i}\right)  + \delta^{jl}\sqrt{\Cal M c^2 + \sum_{p=1}^3(\dot{q}^p)^2}\left(
\frac{\partial\phi_4}{\partial q^l}-\frac{\partial\phi_l}
{\partial q^4}\right)
\\
& - \frac{1}{\Cal M c^2} \left(\Cal M c^2 \delta^{jl}\frac{\partial\psi}{\partial
q^l}+\dot{q}^j\dot{q}^l\,\frac{\partial\psi}{\partial
q^l}+\dot{q}^j\,\sqrt{\Cal M c^2 + \sum_{p=1}^3(\dot{q}^p)^2}
\, \frac{\partial\psi}{\partial q^4}\,\right).
\end{aligned}
\end{equation}

\begin{prop} \textbf{\em(Equations of motion: $\Cal M \ne 0$, four-dimensional observer.)}
Let $g$ be the Minkowski metric, $\phi$ a covariant vector field, and $\psi$ a function on $\R^4$. Extremals of a Lagrangian system defined by the Lagrangian 
\begin{equation}
L = - \sfrac{1}{2} g(\hat u,\hat u) + \phi(\hat u) - \psi =  - \sfrac{1}{2}\Bigl((\dot{q}^4)^2-\sum_{p=1}^3(\dot{q}^p)^2
\Bigr) +\dot{q}^\sigma\phi_\sigma-\psi
\end{equation}
on 
$\R \times T \R^4$, and subject to the constraint
\begin{equation}
g(\hat u,\hat u) = \Cal M c^2 , \quad  \hat u^4 > 0,  \ \Cal M \ne 0,
 \end{equation}
are curves $\gamma(s) = (s, q^\sigma(s))$, satisfying the following system of mixed second and first order differential equations:
\begin{equation} \label{4dimM}
\begin{aligned}
\ddot q^j &=
\dot{q}^l\left(\frac
{\partial\phi_l}{\partial q^j}-\frac{\partial\phi_j}{\partial q^l}\right)  
+ \sqrt{\Cal M c^2 + \sum_{p=1}^3(\dot{q}^p)^2}\left(
\frac{\partial\phi_4}{\partial q^j}-\frac{\partial\phi_j}
{\partial q^4}\right)
\\
& - \frac{\partial\psi}{\partial q^j} - \frac{\dot q^j}{\Cal M c^2} \left(\dot{q}^l\,\frac{\partial\psi}{\partial
q^l} + \sqrt{\Cal M c^2 + \sum_{p=1}^3(\dot{q}^p)^2} \, \frac{\partial\psi}{\partial q^4}\,\right), 
\quad 1 \leq j \leq 3
\end{aligned}
\end{equation}
(equations of motion) and
\begin{equation}
 \dot q^4 = \sqrt{\Cal M c^2 + \sum_{p=1}^3(\dot{q}^p)^2}
 \end{equation}
(equation of the constraint).

The dynamics proceed in the evolution space $\Cal Q _{+} \subset \R \times T\R^4$
defined by the above constraint equation.
\end{prop}

Choosing appropriate coordinates, we can express the above equations in a form adapted to a three-dimensional observer.

Let us denote 
\begin{equation} \label{tcoord}
q^4= ct
\end{equation} 
(time coordinate), and consider on
$J^1\pi \backslash \Cal H_0 \subset  \R \times \R^4 \times \R^4$ new coordinates $(s, q^l, t, v^l, {\dot q}^4)$, defined by the transformation rule
\begin{equation} \label{vcoord}
{\dot q}^l = \frac{1}{c} \, v^l{\dot q}^4, \quad 1 \leq l \leq 3.
\end{equation}
Note that $(s, q^l, t, v^l, {\dot q}^4)$ are {\em global} coordinates, however, no longer fibred coordinates for the original fibration $\pi$.
The meaning of the new coordinates is the following:
$(t, q^l, v^l)$ are coordinates on $\R \times \R^3 \times \R^3$, adapted to the fibration 
${\R}\times{\R}^3 \to {\R}$ of the manifold ${\R}^4$, the fibre of our fibred manifold
$\pi: {\R}\times{\R}^4 \to {\R}$; note that $(q^1,q^2,q^3)$ are cartesian coordinates on $\R^3$. In vector notations $\vec r = (q^1,q^2,q^3)$, and $\vec v = (v^1, v^2, v^3)$ is the usual velocity.

In these coordinates the constraint $\Cal Q$ is given by equation
\begin{equation}
(1 -  \frac{v^2}{c^2}) ({\dot q}^4)^2 = \Cal M c^2. 
\end{equation}
Since we assume $\Cal M \ne 0$, we obtain the evolution space $\Cal Q_{+}$ expressed by equation
\begin{equation}
{\dot q}^4 = \sqrt{\frac{\Cal M c^2}{1 - \frac{v^2}{c^2}}},
\end{equation}
where $v=\sqrt{\sum_{p=1}^3 (v^p)^2} = \sqrt{\sum_{p=1}^3({\rm d}q^p/ {\rm d}t)^2}$ is the usual three-dimensional speed. 
We can also see that along every constrained path,
\begin{equation}
\frac{{\rm d}t}{{\rm d}s} = \frac{1}{c} \dot q^4 = \sqrt{\frac{\Cal M}{1 - \frac{v^2}{c^2}}}.
\end{equation}

Now, equations of motion (\ref{4dimM}) 
can be transformed eliminating the parameter $s$ as follows: First, we have for $j = 1,2,3$
\begin{equation}
\label{acc}
\ddot{q}^j=\frac{{\rm d}}{{\rm d}s}(\dot{q}^j) = \frac{1}{c} \, \frac{{\rm d}}{{\rm d}s} (v^j{\dot q}^4)
= \frac{1}{c} \, \frac{{\rm d}t}{{\rm d}s} \, \frac{{\rm d}}{{\rm d}t} (v^j{\dot q}^4) 
= \sqrt{\frac{\Cal M}{1 - \frac{v^2}{c^2}}} \, \frac{{\rm d}}{{\rm
d}t}\left(v^j\sqrt{\frac{\Cal M}{1 - \frac{v^2}{c^2}}}\right).
\end{equation}
Next, if we denote 
\begin{equation} \label{4pot}
\phi = \phi_\sigma {\rm d}q^\sigma = \phi_l {\rm d}q^l + \phi_4 c\, {\rm d}t  
=  \frac{e}{c} \vec A \, {\rm d} \vec r - eV {\rm d}t,
\end{equation}
i.e.
\begin{equation}
(\phi_l)_{l=1,2,3}= \frac{e}{c}\, \vec{A}, \quad \phi_4= - \frac{e}{c} \, V,
\end{equation} 
where $e$ is
the particle charge, we get from (\ref{force4}) the force
\begin{equation} \label{force3}
\begin{aligned}
\vec{\Cal F} =
e \, \sqrt{\frac{\Cal M}{1 - \frac{v^2}{c^2}}} \, \Bigl( \frac{1}{c} \, \vec{v} \times \rot \vec{A}
- \frac{1}{c} \frac{\partial\vec{A}}{\partial t} - \grad V \Bigr) 
- \frac{\vec{v}}{c^2 - v^2}\frac{{\rm d}\psi}{{\rm d}t} - \grad \psi.
\end{aligned}
\end{equation}

\begin{prop} \textbf{\em(Equations of motion: $\Cal M \ne 0$, three-dimensional observer.)}
Let $\phi$ $(\ref{4pot})$ be a one-form (covariant four-vector field), and $\psi$ a function on the Minkowski space-time $\R^4$. Consider on $(\R \times T\R^4) \backslash \Cal H_0$ adapted coordinates 
$(s,q^l,t,v^l, \dot q^4)$ defined by $(\ref{tcoord})$ and $(\ref{vcoord})$.
Extremals of a Lagrangian system defined by the Lagrangian $\Lambda = L {\rm d}s$,
\begin{equation}
L = - \frac{1}{2} \Bigl(1 - \frac{v^2}{c^2} \Bigr) (\dot{q}^4)^2
+ \frac{1}{c} \Bigl(\frac{e}{c} \vec{A} \vec{v} - eV \Bigr) \dot q^4 - \psi,
\end{equation}
and subject to the constraint
\begin{equation} \label{eqconM}
{\dot q}^4 = \sqrt{\frac{\Cal M c^2}{1 - \frac{v^2}{c^2}}},
\end{equation}
are curves $\c(s) = (t(s),q^l(t(s)))$ in $\R^4$, satisfying the following system of differential equations:
\begin{equation} \label{3dimM}
\begin{aligned}
 \frac{{\rm d}}{{\rm
d}t}\left(\vec{v} \, \sqrt{\frac{\Cal M}{1 - \frac{v^2}{c^2}}}\right)=
e \, \Bigl( \frac{1}{c} \, \vec{v} \times \rot \vec{A}
- \frac{1}{c} \frac{\partial\vec{A}}{\partial t} - \grad V \Bigr) 
- \sqrt{\frac{1- \frac{v^2}{c^2}}{\Cal M}} \Bigl({\rm grad}\,\psi + \frac{\vec{v}}{c^2 - v^2}
\frac{{\rm d}\psi}{{\rm d} t} \Bigr)
\end{aligned}
\end{equation}
(equations of motion) and
\begin{equation} \label{masseq}
\frac{{\rm d}t}{{\rm d}s} = \sqrt{\frac{\Cal M}{1 - \frac{v^2}{c^2}}}
\end{equation}
(equation of the constraint).
The dynamics proceed in the evolution space $\Cal Q _{+} \subset \R \times T\R^4$
defined by equation $(\ref{eqconM})$.
\end{prop}

We shall see later that equation of the constraint (\ref{masseq}) has the meaning of {\em mass equation}, or, if multiplied by $c^2$, of {\em energy equation}.

In what follows, we shall denote by $\vec{\Cal F}_L$ the Lorentz force,
\begin{equation}
\vec{\Cal F}_L = e \, \Bigl( \frac{1}{c} \, \vec{v} \times \rot \vec{A}
- \frac{1}{c} \frac{\partial\vec{A}}{\partial t} - \grad V \Bigr).
\end{equation}

\subsection{The constraint force}

Chetaev equations of motion contain the constraint force $\Phi$. It is defined on 
$J^1 \pi = \R \times T\R^4$ and depends upon one Lagrange multiplier $\lambda$ (to be determined). For 
\begin{equation}
f = \dot q^4 - \sqrt{\Cal M c^2 + \sum_{p=1}^3(\dot{q}^p)^2} 
\end{equation} 
it reads
\begin{equation} \label{cofo}
\Phi = - \sum_{l=1}^3\frac{\lambda \dot{q}^l}{\sqrt{\Cal M c^2+\sum_{p=1}^3
(\dot{q}^p)^2}}\, {\rm d}q^l\wedge{\rm d}s + \lambda \, {\rm d}q^4\wedge{\rm d}s.
\end{equation}
Chetaev equations take the form
\begin{equation} \label{def-rovnice-1}
\begin{aligned}
-\ddot{q}^l+\dot{q}^\sigma\left(\frac{\partial\phi_
\sigma}{\partial q^l}-\frac{\partial\phi_l}{\partial
q^\sigma}\right) - \frac{\partial\psi}{\partial q^l} &=
\frac{- \lambda \dot{q}^l}{\sqrt{\Cal M c^2 + \sum_{p=1}^3 (\dot{q}^p)^2}},
\\
\ddot{q}^4+\dot{q}^\sigma\left(\frac{\partial\phi_\sigma}
{\partial q^4}-\frac{\partial\phi_4}{\partial q^\sigma}\right) -
\frac{\partial \psi}{\partial q^4} &= \lambda.
\end{aligned}
\end{equation}
Multiplying the $\nu$-th of the above $4$ equations by $\dot{q}^\nu$ and summing over 
$\nu = 1,2,3,4$, we obtain
\begin{equation} \label{cons}
(\ddot{q}^4\dot{q}^4-\sum_{l=1}^3\ddot{q}^l\dot{q}^l) + \dot{q}^\sigma
\dot{q}^\nu\left(\frac{\partial\phi_\sigma}{\partial q^\nu} -
\frac{\partial\phi_\nu}{\partial
q^\sigma}\right) - \dot{q}^\nu \frac{\partial\psi}{\partial q^\nu} = 
\lambda \,  \frac{\Cal M c^2}{\sqrt{\Cal M c^2+\sum_{p=1}^3(\dot{q}^p)^2}}.
\end{equation}
Solutions $\c(s) = (q^\sigma(s))$ of Chetaev equations satisfy the equation of the constraint, and consequently, also its derivative,
\begin{equation}
\ddot q^4 = \frac{\frac{1}{2} \frac{{\rm d}}{{\rm d}s} \Cal M c^2 + \sum_{l=1}^3 \ddot q^l \dot{q}^l}{\sqrt{\Cal M c^2+\sum_{p=1}^3 (\dot{q}^p)^2}},
\quad \text{and then} \quad \ddot q^4 \dot q^4 =  \frac{1}{2} \, \frac{{\rm d}}{{\rm d}s} \Cal M c^2 + \sum_{l=1}^3 \ddot q^l \dot{q}^l.
\end{equation}
Hence, the first term in brackets on the left-hand side of equation (\ref{cons})
is $\frac{{\rm d}}{{\rm d}s}(\frac{1}{2} \Cal M c^2)$, the second term is identically zero because of skewsymmetry of the expression in brackets, and the third term equals $- {\rm d} \psi / {\rm d}s$.
In this way, along solutions of Chetaev equations passing in the evolution space $\Cal Q_{+}$ we have
\begin{equation} \label{psider}
\frac{{\rm d}}{{\rm d}s} \Bigl(\sfrac{1}{2} \Cal M c^2 - \psi\Bigr) =
\lambda \, \frac{\Cal M c^2}{\sqrt {\Cal M c^2 + \sum_{p=1}^3(\dot{q}^p)^2}}.
\end{equation}
If $\Cal M \ne 0$, we can compute the multiplier $\lambda$:
\begin{equation}
\lambda = \frac{\sqrt {\Cal M c^2 + \sum_{p=1}^3(\dot{q}^p)^2}}{\Cal M c^2}
\, \frac{{\rm d}}{{\rm d}s} \Bigl(\sfrac{1}{2} \Cal M c^2 - \psi\Bigr).
\end{equation}
Writing $\Phi = \hat \Phi_\sigma {\rm d}q^\sigma \land {\rm d}s$ and substituting the obtained expression for $\lambda$ into formula (\ref{cofo}) for $\Phi$, we get {\em components of the constraint force along solutions of the constraint equations} as follows:
\begin{equation}
\begin{aligned}
\hat \Phi_l &=  - \frac{\dot q^l}{\Cal M c^2}
\, \frac{{\rm d}}{{\rm d}s} \left(\sfrac{1}{2} \Cal M c^2 - \psi\right)
= \frac{\dot q^l}{\Cal M c^2}
\, \frac{{\rm d}}{{\rm d}s} \left(\psi - \sfrac{1}{2} \Cal M c^2 \right),
\\
\hat \Phi_4 &= \frac{\sqrt {\Cal M c^2 + \sum_{p=1}^3(\dot{q}^p)^2}}{\Cal M c^2}
\, \frac{{\rm d}}{{\rm d}s} \left(\sfrac{1}{2} \Cal M c^2 - \psi\right)
\end{aligned}
\end{equation}
in coordinates $(s,q^\sigma,\dot q^\sigma)$, and
\begin{equation}
\begin{aligned}
\Phi_l &= \hat \Phi_l
= \frac{v^l}{ c^2 - v^2} \, \frac{{\rm d}}{{\rm d}t} \left(\psi - \sfrac{1}{2} \Cal M c^2 \right),
\\
\Phi_4 &= c \, \hat \Phi_4
= - \, \frac{c^2}{c^2 - v^2} \, \frac{{\rm d}}{{\rm d}t} \Bigl(\psi - \sfrac{1}{2} \Cal M c^2 \Bigr)
\end{aligned}
\end{equation}
in coordinates $(s,q^l,t,v^l,\dot q^4)$.

For the sake of simplicity we introduce the following vector notation, that takes into account the formula for the constraint force and the relation between equations (\ref{4dimM}) and (\ref{3dimM})
(note that they are connected via a multiplier)
\begin{equation} 
{\vec{\Cal F}}_C = - \sqrt{\frac{1-\frac{v^2}{c^2}}{\Cal M}}\, (\Phi_l)_{1 \leq  l \leq 3} = 
-  \frac{\vec v}{c^2 \sqrt{\Cal M (1-\frac{v^2}{c^2})}} \,\frac{\rm d}{{\rm d}t}\left( \psi - \sfrac{1}{2} \Cal M c^2 \right),
\end{equation}
and call ${\vec{\Cal F}}_C$ the {\em induced constraint force}.
Note that ${\vec{\Cal F}}_C$ is defined on the evolution space $\Cal Q_{+}$, and vanishes
whenever $\vec v = 0$ or
\begin{equation}
\psi - \sfrac{1}{2} \Cal M c^2 = \text{constant of the motion}.
\end{equation}

The constraint force appears in the motion equations (\ref{3dimM}):

\begin{prop}
If $\Cal M \ne 0$ then equations of motion (\ref{3dimM}) can be expressed as follows:
\begin{equation}
 \frac{{\rm d}}{{\rm d}t}\Bigl(\vec{v} \, \sqrt{\frac{\Cal M}{1 - \frac{v^2}{c^2}}}\Bigr)=
{\vec{\Cal F}}_L + {\vec{\Cal F}}_C
- \sqrt{\frac{1- \frac{v^2}{c^2}}{\Cal M}} {\rm grad}\,\psi  - \frac{1}{2} \frac{\vec v}{\sqrt{\Cal M (1-\frac{v^2}{c^2})}} \frac{{\rm d} \Cal M}{{\rm d} t},
\end{equation}
where ${\vec{\Cal F}}_L$ is the Lorentz force and ${\vec{\Cal F}}_C$ is the induced constraint force.
\end{prop}

In the rest of the paper we shall be interested in the case $\Cal M = \const.$ We shall discuss
separately the cases $\Cal M > 0$, $\Cal M < 0$, and $\Cal M = 0$. 
Note for constant $\Cal M$ the above equation takes the form
\begin{equation} \label{eqmoconst}
 \frac{{\rm d}}{{\rm d}t}\Bigl(\vec{v} \, \sqrt{\frac{\Cal M}{1 - \frac{v^2}{c^2}}}\Bigr)=
{\vec{\Cal F}}_L + {\vec{\Cal F}}_C
- \sqrt{\frac{1- \frac{v^2}{c^2}}{\Cal M}} {\rm grad}\,\psi.
\end{equation}


\section{Particles with positive square of mass}

Let us suppose {\em $\Cal M > 0$ and constant on $\R^4$}. We set
\begin{equation}
\Cal M = m_0^2.
\end{equation}

As above, we consider either fibred coordinates $(s, q^\sigma, \dot q^\sigma)$, $1 \leq \sigma \leq 4$, or adapted coordinates $(s,q^l,t,v^l,\dot q^4)$, $1 \leq l \leq 3$, on $(\R \times T\R^4) \backslash \Cal H_0$.

\subsection{The evolution space}

The constraint $\Cal Q \subset (\R \times T\R^4)  \backslash \Cal H_0$ is given by equation
\begin{equation}
(\dot q^4)^2 - \sum_{p=1}^3(\dot{q}^p)^2 = m_0^2 c^2, \quad \text{resp.} \quad
(\dot q^4)^2 \left(1-\frac{v^2}{c^2} \right) = m_0^2 c^2.
\end{equation}
The latter implies $c^2 - v^2 > 0$, i.e.
\begin{equation}
v < c.
\end{equation}
Thus, {\em the speed of a relativistic particle with $\Cal M  > 0$ is lower than the light speed}. 

For the evolution space $\Cal Q_+$ we then have
\begin{equation}
\label{Q+M+}
\dot q^4 = \frac{m_0 c}{\sqrt{1-\frac{v^2}{c^2}}},
\end{equation}
where $m_0 > 0$. We can see that the constant $m_0$ has the meaning of the {\em rest mass} of
a particle.

We also denote, as usual, mass and (kinetic) energy by
\begin{equation}
m = \frac{m_0}{\sqrt{1-\frac{v^2}{c^2}}}, \quad 
\Cal E = m c^2.
\end{equation}

\smallskip

We can ``visualise" the constraint $\Cal Q$ and the evolution space $\Cal Q_+$ in the ``velocity space" as follows: For illustration we supress one dimension by considering $\dot q^3 = 0$ and denote
$(\dot{q}^1,\dot{q}^2,\dot{q}^4)= c (x,y,w)$. Then $\Cal Q$ is given by
\begin{equation}
\frac{w^2}{m_0^2}-\frac{x^2}{m_0^2}-\frac{y^2}{m_0^2}=1
\end{equation}
(Fig. 1).
\begin{figure}
\centering
\includegraphics[width=8cm]{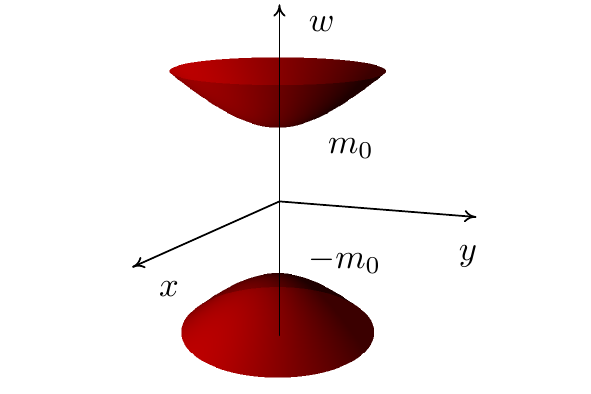}
\caption
{Projection of the evolution space for $\Cal M > 0$.}
\end{figure}

\subsection{Four-momentum}

Since $\R^4$ is equipped with the Minkowski metric field $g$ which is a regular symmetric 
$(2,0)$-tensor field on $\R^4$, every contravariant four-vector field $\hat u$ on $\R^4$ is canonically associated with a covariant four-vector field (a $1$-form) $g(\hat u, \cdot)$ on $\R^4$. We put
\begin{equation}
p = g(\hat u, \cdot)
\end{equation}
and call $p$ the {\em four-momentum} associated with $\hat u$. In fibred coordinates, where the section
$\hat u$ has components $\hat u^\sigma = \dot q^\sigma \circ \hat u$, $1 \leq \sigma \leq 4$, we have 
$p = \hat p_\sigma {\rm d}q^\sigma$, where
\begin{equation}
\hat p_\sigma = g_{\sigma\nu} u^\nu = g_{\sigma \nu} (\dot q^\sigma \circ \hat u),
\quad \text{i.e.} \quad
(\hat p_\sigma)_{1 \leq \sigma \leq 4} = (- \dot q^1, - \dot q^2, - \dot q^3, \dot q^4) \circ \hat u.
\end{equation}
For simplicity of notations, we shall denote components of $p$ associated with $\hat u$ simply by
\begin{equation}
(\hat p_\sigma)_{1 \leq \sigma \leq 4} = (- \dot q^1, - \dot q^2, - \dot q^3, \dot q^4),
\end{equation}
where the $\dot q$'s are components of $\hat u$.

In adapted coordinates $(s,q^l,t,v^l,\dot q^4)$ we similarly have $p = p_l {\rm d}q^l + p_4 {\rm d}t$,
\begin{equation}
p_l = - \frac{v^l \dot q^4}{c}, \quad p_4 = c \dot q^4.
\end{equation}
Note that, in the latter case, the corresponding ``contravariant components" are 
\begin{equation}
p^l = \frac{v^l \dot q^4}{c}, \quad p^4 = c \dot q^4.
\end{equation}
We denote ${\vec p} = (p^l)_{1 \leq l \leq 3}$, and write
\begin{equation}
(p_\sigma)_{1 \leq \sigma \leq 4} = (- \frac{{\vec v} \dot q^4}{c}, \, c \dot q^4),
\quad \
(p^\sigma)_{1 \leq \sigma \leq 4} = (\frac{{\vec v} \dot q^4}{c}, \, c \dot q^4).
\end{equation}

With help of four-momentum, {\em equation of the constraint $\Cal Q$}, i.e. $g(\hat u,\hat u) = m_0^2 c^2$, reads
\begin{equation}
\hat p_\sigma \hat p^\sigma = m_0^2 c^2, \quad \text{or} \quad
p_l p^l + \frac{1}{c^2} p_4 p^4 = m_0^2 c^2.
\end{equation}

On the evolution space $\Cal Q_+$, where
\begin{equation} \label{pcon}
\dot q^4 = \frac{m_0 c}{\sqrt{1 - \frac{v^2}{c^2}}} = mc = \frac{\Cal E}{c} \, ,
\end{equation}
components of the four-momentum take the form
\begin{equation}
(\hat p_\sigma)_{1 \leq \sigma \leq 4} = (- m{\vec v}, \, \Cal E / c),
\quad \
(\hat p^\sigma)_{1 \leq \sigma \leq 4} = (m {\vec v}, \, \Cal E / c),
\end{equation}
Finally, with help of (\ref{pcon}) we can compute the energy; we obtain the familiar formula
\begin{equation}
\Cal E = c \sqrt{m_0^2 c^2 - p_l p^l} = c \sqrt{m_0^2 c^2 + p^2},
\end{equation}
where $p^2 = \vec p \cdot \vec p = \sum_{l=1}^{3} (p^l)^2$ denotes the usual square of length of the three-vector $\vec p$.

\subsection{Equations of motion}

Rewriting motion equations for the particular case $\Cal M = m_0^2 = \const.$, we obtain: 

\begin{prop} \textbf{\em(Equations of motion: usual particles, four-dimensional observer.)}
Let $g$ be the Minkowski metric, $\phi$ a covariant vector field, and $\psi$ a function on $\R^4$. Extremals of a Lagrangian system defined by the Lagrangian $\Lambda = L {\rm d}s$,
\begin{equation}
L = - \sfrac{1}{2} g(\hat u,\hat u) + \phi(\hat u) - \psi =  - \sfrac{1}{2}\Bigl((\dot{q}^4)^2-\sum_{p=1}^3(\dot{q}^p)^2 \Bigr) +\dot{q}^\sigma\phi_\sigma-\psi,
\end{equation}
on 
$\R \times T \R^4$, and subject to the constraint
\begin{equation}
\dot q^4 = \sqrt{m_0^2 c^2 + \sum_{l=1}^3(\dot{q}^l)^2},
\end{equation}
are curves $\gamma(s) = (s, q^\sigma(s))$, satisfying the system of mixed second and first order differential equations as follows:  (\ref{4dimM}) where $\Cal M = m_0^2$ (equations of motion), and
$ c\dot q^4 = \Cal E$ (equation of the constraint, energy equation).
\end{prop}

\begin{prop} \label{3dim-usual}
\textbf{\em(Equations of motion: usual particles, three-dimensional observer.)}
Let $\phi$ $(\ref{4pot})$ be a one-form (covariant four-vector field), and $\psi$ a function on the Minkowski space-time $\R^4$. Consider on $(\R \times T\R^4) \backslash \Cal H_0$ adapted coordinates 
$(s,q^l,t,v^l, \dot q^4)$ defined by $(\ref{tcoord})$ and $(\ref{vcoord})$.
Extremals of a Lagrangian system defined by the Lagrangian $\Lambda = L {\rm d}s$,
\begin{equation}
L = - \frac{1}{2} \Bigl(1 - \frac{v^2}{c^2} \Bigr) (\dot{q}^4)^2
+ e \, \Bigl(\frac{1}{c} \vec{A} \vec{v} - V \Bigr) \frac{\dot q^4}{c}- \psi,
\end{equation}
and subject to the constraint
\begin{equation} 
{\dot q}^4 = \frac{m_0 c}{\sqrt{1 - \frac{v^2}{c^2}}},
\end{equation}
are curves $\c(s) = (t(s),q^l(t(s)))$ in $\R^4$, satisfying the following system of differential equations:
\begin{equation} \label{eqm+3}
 \frac{{\rm d}}{{\rm d}t} \Bigl({\frac{m_0 \vec v}{\sqrt{1 - \frac{v^2}{c^2}}}} \Bigr)
 = {\vec{\Cal F}}_L + {\vec{\Cal F}}_C
- \frac{1}{m_0} \, \sqrt{1- \frac{v^2}{c^2}} \ {\rm grad}\,\psi
\end{equation}
(equations of motion), together with the following equation of the constraint (mass equation, resp. energy equation):
\begin{equation} \label{meq}
\frac{{\rm d}t}{{\rm d}s} = m, \quad \text{resp.} \  \quad \
c^2\frac{{\rm d}t}{{\rm d}s} = \Cal E.
\end{equation}
\end{prop}

Note that the constraint force along solutions of the equations of motion takes the form
\begin{equation} 
{\vec{\Cal F}}_C = 
- \frac{1}{m_0 c^2} \, \frac{\vec v}{\sqrt{1 - \frac{v^2}{c^2}}} \, \frac{{\rm d \psi}}{{\rm d} t}.
\end{equation}
Hence, along solutions, ${\vec{\Cal F}}_C = 0$ if $\vec v = 0$ or {\em $\psi$ is a constant of the motion}.

\subsection{Variational principle for time-dependent curves} \label{secvar}

Taking into account equations of motion (\ref{eqm+3}), we can ask if they are equations for extremals of an (unconstrained) variational principle for curves 
$t \to (q^1(t), q^2(t), q^3(t))$ in $\R^3$.
This means, we ask about a {\em Lagrangian $1$-form $\Cal L(t, q^l, v^l) {\rm d}t$ on 
$\R \times \R^3$} such that equations (\ref{eqm+3}) would identify with Euler--Lagrange equations
\begin{equation}
\frac{\pr \Cal L}{\pr q^l} - \frac{\rm d}{{\rm d}t} \frac{\pr \Cal L}{\pr v^l} = 0.
\end{equation}
The problem can be tackled directly by applying Helmholtz conditions (necessary and sufficient conditions for a system of differential equations be variational ``as it stands") \cite{Hel}. However, the answer is known: as shown in
\cite{Nov}, equations with the left-hand side as in (\ref{eqm+3}) are variational if the force on the right-hand side is a Lorentz-type force. Hence, we can conclude:

\begin{prop}
If the scalar field $\psi$ is a constant then equations (\ref{eqm+3}) are variational as equations for extremals of the Lagrangian
$\Cal L(t, q^l, v^l) {\rm d}t$ on $\R \times \R^3$, such that
\begin{equation}
\Cal L = - m_0c^2 \sqrt{1 - \frac{v^2}{c^2}} + \frac{e}{c} \vec{A} \vec{v} - eV.
\end{equation}
\end{prop}

Note an interesting connection between {\em vanishing of  the induced constraint force and the existence of a Lagrangian}, namely $\vec{\Cal F}_C \equiv 0$ implies that ${\rm d}\psi/{\rm d}t = 0$, i.e. 
$\psi = \const$.

\begin{rem}
Within the theory of non-holonomic systems it is known that motion equations of a Lagrangian system subject to constraints need not be Euler--Lagrange equations of the constrained Lagrangian. This is the case also in our situation. Indeed, the unconstrained Lagrangian $1$-form
\begin{equation}
\Lambda = L {\rm d}s = - \Bigl( \frac{1}{2} \Bigl(1 - \frac{v^2}{c^2} \Bigr) (\dot{q}^4)^2
- \Bigl(\frac{e}{c} \vec{A} \vec{v} - eV \Bigr) \frac{\dot q^4}{c} + \psi \Bigr) {\rm d}s
\end{equation}
gives rise to the constrained Lagrangian
\begin{equation}
\begin{aligned}
\Lambda_C &=  L_C {\rm d}s = - \Bigl( \frac{1}{2} m_0c^2 \sqrt{1 - \frac{v^2}{c^2}} 
- \frac{e}{c} \vec{A} \vec{v} + eV + \frac{\psi}{m}  \Bigr) {\rm d}t
\\
&= - \Bigl( m_0c^2 \sqrt{1 - \frac{v^2}{c^2}} - \frac{e}{c} \vec{A} \vec{v} + eV + \frac{\psi}{m}  \Bigr) {\rm d}t 
+ \frac{1}{2} m_0^2c^2 {\rm d}s 
\end{aligned}
\end{equation}
defined on the constraint $\Cal Q_+$.  The relevant part of the Lagrangian $\Lambda_C$ for considering time-parametrised curves is its horizontal part with respect to the projection onto the time axis, i.e. the $1$-form $\Cal L {\rm d}t$, with
\begin{equation}
\Cal L = - m_0 c^2 \sqrt{1 - \frac{v^2}{c^2}} + \frac{e}{c} \vec{A} \vec{v} - eV - \frac{\psi}{m}. 
\end{equation}
However, Euler--Lagrange equations of $\Cal L$ do not coincide with equations (\ref{eqm+3}) unless 
$\psi$ is a constant.
\end{rem}

\subsection{Dicke force}

Equations of motion in proposition \ref{3dim-usual} can be equivalently expressed in a way that
admits to compare the obtained forces  with a hypothesis due to Dicke on non-Lorentz type interactions \cite{dicke}.

Denote $\exp \mu = e^\mu$ and put
\begin{equation} \label{Dickefield}
\mu =  \frac{\psi}{m_0^2 c^2}, \quad \tilde m_0 = m_0 \, e^\mu.
\end{equation}
Let us express motion equations (\ref{eqm+3}) in terms of the new scalar potential 
$\mu$ and ``mass" $\tilde m_0$. We get
\begin{equation} \label{D1}
 \frac{{\rm d}}{{\rm d}t}\Bigl(\frac{\tilde m_0 e^{- \mu} \vec{v}}{\sqrt{1 - \frac{v^2}{c^2}}} \Bigr) 
 + \frac{m_0 \vec v}{\sqrt{1 - \frac{v^2}{c^2}}} \, \frac{{\rm d \mu}}{{\rm d} t} =
\vec{\cal{F}}_L -  m_0 c^2 \,\sqrt{1- \frac{v^2}{c^2}}  \, \grad \mu,
\end{equation}
and
\begin{equation}
 \frac{{\rm d}e^{-\mu}}{{\rm d} t} = - e^{-\mu}  \frac{{\rm d \mu}}{{\rm d} t},
\end{equation}
hence, 
\begin{equation}
e^{-\mu} \frac{{\rm d}}{{\rm d}t}\Bigl(\frac{\tilde m_0 \vec{v}}{\sqrt{1 - \frac{v^2}{c^2}}} \Bigr) 
= \vec{\cal{F}}_L -  m_0 c^2 \,\sqrt{1- \frac{v^2}{c^2}}  \, \grad \mu.
\end{equation}
Multiplying this equation by $e^\mu$ we have
\begin{equation}
\frac{{\rm d}}{{\rm d}t}\Bigl(\frac{\tilde m_0 \vec{v}}{\sqrt{1 - \frac{v^2}{c^2}}} \Bigr) 
= e^\mu \, \vec{\cal{F}}_L -  \tilde m_0 c^2 \,\sqrt{1- \frac{v^2}{c^2}}  \, \grad \mu.
\end{equation}
However,
\begin{equation}
\grad \tilde m_0 = m_0 \grad e^\mu = m_0 e^\mu \grad \mu = \tilde m_0 \grad \mu,
\end{equation}
and we finally obtain
\begin{equation}
\frac{{\rm d}}{{\rm d}t}\Bigl(\frac{\tilde m_0 \vec{v}}{\sqrt{1 - \frac{v^2}{c^2}}} \Bigr) 
= e^\mu \, \vec{\cal{F}}_L -  c^2 \,\sqrt{1- \frac{v^2}{c^2}}  \, \grad \tilde m_0.
\end{equation}

\begin{prop} \label{3dim-Dicke}
With help of the mass function $\tilde m_0$, depending upon a scalar potential $\psi$ on $\R^4$, and defined by 
\begin{equation}
 \tilde m_0 = m_0 \exp \left(\frac{\psi}{m_0^2 c^2}\right),
\end{equation}
equations of motion in proposition \ref{3dim-usual} take the following equivalent form:

$\bullet$ in absence of the electromagnetic field:
\begin{equation}
\frac{{\rm d}}{{\rm d}t}\Bigl(\frac{\tilde m_0 \vec{v}}{\sqrt{1 - \frac{v^2}{c^2}}} \Bigr) 
= -  c^2 \,\sqrt{1- \frac{v^2}{c^2}}  \, \grad \tilde m_0,
\end{equation}

$\bullet$ in presence of the electromagnetic field:
\begin{equation}
\frac{{\rm d}}{{\rm d}t}\Bigl(\frac{\tilde m_0 \vec{v}}{\sqrt{1 - \frac{v^2}{c^2}}} \Bigr) 
= \frac{\tilde m_0}{m_0} \, \vec{\cal{F}}_L -  c^2 \,\sqrt{1- \frac{v^2}{c^2}}  \, \grad \tilde m_0.
\end{equation}

Hence, the particle moves like having a non-constant rest mass and subject to a force depending upon the particle speed and the gradient of the mass. The influence of an electromagnetic field on the particle's dynamics depends not only upon the particle charge, but also upon 
$\exp \left(\psi / (m_0^2 c^2)\right) = \tilde m_0 / m_0$.
\end{prop}

Let us denote 
\begin{equation}
{\vec {\Cal F}}_D = -  c^2 \,\sqrt{1- \frac{v^2}{c^2}}  \, \grad \tilde m_0
\end{equation}
and call ${\vec {\Cal F}}_D$ {\em Dicke force}. 

\begin{rem}
The obtained formula for the force $\vec{\Cal F}_D$ complies with a prediction of a relativistic non-Lorentz type interaction due to Dicke $\cite{dicke}$. Although in his paper a formula or, at least, exact arguments are missing, Dicke conjectured the existence of a force, originating from the mass distribution of the universe, and acting on particles with nonzero (positive) rest mass. By his hypothesis, the particle should move like having a nonconstant mass depending upon a scalar field, and subject to a force depending upon the particle's nonconstant mass, and upon its speed proportionally to the factor $- \sqrt{1- v^2/c^2}$. Remarkably, this conjecture
was influenced by Brans and Dicke's modification of general relativity, published earlier, attempting to adapt the general relativity theory to the Mach principle (see $\cite{bransdicke}$).
\end{rem}


\section{Tachyons}

Let us suppose {\em $\Cal M < 0$ and constant on $\R^4$}. We set
\begin{equation}
\Cal M = - m_0^2,
\end{equation}
where $m_0 > 0$. Particles of this kind are {\em particles with negative square of mass, called tachyons}.

Again we consider either fibred coordinates $(s, q^\sigma, \dot q^\sigma)$, $1 \leq \sigma \leq 4$, or adapted coordinates $(s,q^l,t,v^l,\dot q^4)$, $1 \leq l \leq 3$, on $(\R \times T\R^4) \backslash \Cal H_0$.

\subsection{The evolution space}

The constraint $\Cal Q \subset (\R \times T\R^4) \backslash \Cal H_0$ is given by equation
\begin{equation}
(\dot q^4)^2 - \sum_{p=1}^3(\dot{q}^p)^2 = - m_0^2 c^2, \quad \text{resp.} \quad
(\dot q^4)^2 \left(1-\frac{v^2}{c^2} \right) = - m_0^2 c^2.
\end{equation}
The evolution space $\Cal Q_+$ is then the submanifold
\begin{equation}
\label{Q+M+}
\dot q^4 = \sqrt{\frac{- m_0^2 c^2}{1- \frac{v^2}{c^2}}}.
\end{equation}

For tachyons one can introduce {\em instantaneous mass} and {\em (kinetic) energy} by the following formulas
\begin{equation}
m =  \sqrt{\frac{- m_0^2}{1- \frac{v^2}{c^2}}} ,
 \quad \text{i.e.:} \quad
m = \frac{m_0}{\sqrt{\frac{v^2}{c^2} - 1}} \ \text{where $v > c$}, 
  \quad \text{or} 
\quad  m = \frac{im_0}{\sqrt{1 - \frac{v^2}{c^2}}}  \ \text{where $v < c$},
\end{equation}
\begin{equation}
\Cal E = m c^2.
\end{equation}
By the above, {\em tachyons can be regarded either as particles with real positive mass $m_0$ moving with a speed $v > c$, or as particles with imaginary mass $im_0$, moving with speed lower than the light speed (like ``normal particles")}. 

In our setting, when $\dot q^4 \in \R$, tachyons appear as particles with real mass $m_0 > 0$ moving with a speed greater than the light speed. In this case the quantities $m$ and $\Cal E$ are {\em real}. Note that if the speed of a tachyon is increasing, its mass $m$ is decreasing ($v \to \infty$ means that $m \to 0$ and $\Cal E \to 0$, and conversely, $v \to c$ means that $m \to \infty$ and $\Cal E \to \infty$), and that 
\begin{equation}
m = m_0 \quad \text{if} \quad v = c \sqrt{2}.
\end{equation}

Similarly as in the case $\Cal M > 0$ we can ``visualise" the constraint $\Cal Q$ and the evolution space $\Cal Q_+$ in the ``velocity space". Supressing the $\dot q^3$ dimension and denoting again
$(\dot{q}^1,\dot{q}^2,\dot{q}^4)= c (x,y,w)$, the manifold $\Cal Q$ is given by equation
\begin{equation}
\frac{w^2}{m_0^2}-\frac{x^2}{m_0^2}-\frac{y^2}{m_0^2}= -1
\end{equation}
(Fig. 2).
\begin{figure}
\centering
\includegraphics[width=8cm]{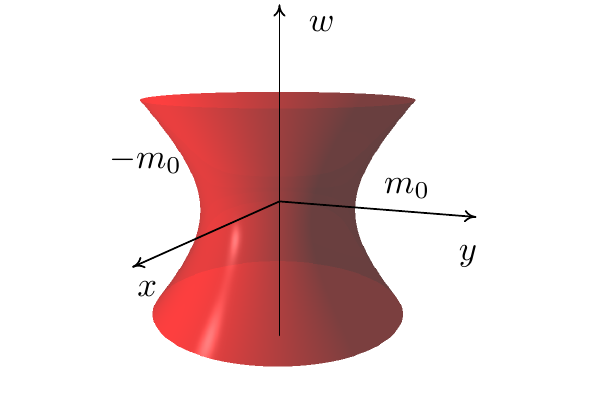}
\caption
{Projection of the evolution space of a tachyon ($\Cal M < 0$).}
\end{figure}

\subsection{Four-momentum and energy}

We can introduce {\em four-momentum} in a full analogy with the $\Cal M > 0$ case setting
\begin{equation}
p = g(\hat u, \cdot).
\end{equation}
In coordinates $p = \hat p_\sigma {\rm d}q^\sigma = p_l {\rm d}q^l + p_4{\rm d}t$, and we obtain the same formulas for the components as in the previous section:
\begin{equation}
\begin{aligned}
(\hat p_\sigma)_{1 \leq \sigma \leq 4}  &= (- \dot q ^1, - \dot q ^2,- \dot q ^3, \dot q ^4),
\\
(p_\sigma)_{1 \leq \sigma \leq 4} &= (- \frac{{\vec v} \dot q^4}{c}, \, c \dot q^4),
\quad \
(p^\sigma)_{1 \leq \sigma \leq 4} = (\frac{{\vec v} \dot q^4}{c}, \, c \dot q^4).
\end{aligned}
\end{equation}

For the {\em equation of the constraint $\Cal Q$}, i.e. $g(\hat u,\hat u) = - m_0^2 c^2$, we obtain
\begin{equation} \label{pcon-}
\hat p_\sigma \hat p^\sigma = - m_0^2 c^2, \quad \text{or} \quad
p_l p^l + \frac{1}{c^2} p_4 p^4 = - m_0^2 c^2.
\end{equation}

On the evolution space $\Cal Q_+$, where
\begin{equation}
\dot q^4 = \frac{m_0 c}{\sqrt{\frac{v^2}{c^2}-1}} = mc = \frac{\Cal E}{c} \, ,
\end{equation}
components of the four-momentum take the form
\begin{equation}
(\hat p_\sigma)_{1 \leq \sigma \leq 4} = (- m{\vec v}, \, \Cal E / c),
\quad \
(\hat p^\sigma)_{1 \leq \sigma \leq 4} = (m {\vec v}, \, \Cal E / c).
\end{equation}
Finally, with help of (\ref{pcon-}) we can compute the energy:
\begin{equation}
\Cal E = c \sqrt{- p_l p^l-m_0^2 c^2 } = c \sqrt{p^2 - m_0^2 c^2},
\end{equation}
where, as in the previous sections, $p^2 = \vec p \cdot \vec p$ is the usual square of length of the three-vector $\vec p$.

\subsection{Equations of motion} 

Rewriting motion equations (\ref{4dimM}) for $\Cal M = - m_0^2 = \const.$, we obtain: 

\begin{prop} \textbf{\em(Equations of motion: tachyons, four-dimensional observer.)}
Let $g$ be the Minkowski metric, $\phi$ a covariant vector field, and $\psi$ a function on $\R^4$. Extremals of a Lagrangian system defined by the Lagrangian $\Lambda = L {\rm d}s$,
\begin{equation}
L = - \sfrac{1}{2} g(\hat u,\hat u) + \phi(\hat u) - \psi =  - \sfrac{1}{2}\Bigl((\dot{q}^4)^2-\sum_{p=1}^3(\dot{q}^p)^2 \Bigr) +\dot{q}^\sigma\phi_\sigma-\psi,
\end{equation}
on 
$\R \times T \R^4$, and subject to the constraint
\begin{equation}
\dot q^4 = \sqrt{\sum_{l=1}^3(\dot{q}^l)^2 - m_0^2 c^2}
\end{equation}
are curves $\gamma(s) = (s, q^\sigma(s))$, satisfying the following system of mixed second and first order differential equations: (\ref{4dimM}) where $\Cal M = - m_0^2$ (equations of motion), and
$ c\dot q^4 = \Cal E$ (equation of the constraint, energy equation).
\end{prop}

\begin{prop} \label{3dim-tach}
\textbf{\em(Equations of motion: tachyons, three-dimensional observer.)}
Let $\phi$ $(\ref{4pot})$ be a one-form (covariant four-vector field), and $\psi$ a function on the Minkowski space-time $\R^4$. Consider on $(\R \times T\R^4) \backslash \Cal H_0$ adapted coordinates 
$(s,q^l,t,v^l, \dot q^4)$ defined by $(\ref{tcoord})$ and $(\ref{vcoord})$.
Extremals of a Lagrangian system defined by the Lagrangian $\Lambda = L {\rm d}s$,
\begin{equation}
L = - \frac{1}{2} \Bigl(1 - \frac{v^2}{c^2} \Bigr) (\dot{q}^4)^2
+ e \, \Bigl(\frac{1}{c} \vec{A} \vec{v} - V \Bigr) \frac{\dot q^4}{c}- \psi,
\end{equation}
and subject to the constraint
\begin{equation} 
{\dot q}^4 = \frac{m_0 c}{\sqrt{\frac{v^2}{c^2} - 1}},
\end{equation}
are curves $\c(s) = (t(s),q^l(t(s)))$ in $\R^4$, satisfying the following system of differential equations:
\begin{equation} \label{eq-m+3}
 \frac{{\rm d}}{{\rm d}t} \Bigl({\frac{m_0 \vec v}{\sqrt{\frac{v^2}{c^2} - 1}}} \Bigr)
 = {\vec{\Cal F}}_L + {\vec{\Cal F}}_C
- \frac{1}{m_0} \, \sqrt{\frac{v^2}{c^2} - 1} \ {\rm grad}\,\psi 
\end{equation}
(equations of motion), together with the following equation of the constraint (mass equation, resp. energy equation):
\begin{equation} \label{-meq}
\frac{{\rm d}t}{{\rm d}s} = m, \quad \text{resp.} \  \quad \
c^2\frac{{\rm d}t}{{\rm d}s} = \Cal E.
\end{equation}
\end{prop}

The constraint force along solutions of the equations of motion in this case  takes the form
\begin{equation}
{\vec{\Cal F}}_C =  \,\frac{1}{m_0 c^2} \, \frac{\vec v}{\sqrt{\frac{v^2}{c^2} - 1}} \, \frac{{\rm d \psi}}{{\rm d} t}.
\end{equation}
Along solutions, ${\vec{\Cal F}}_C = 0$ if $\vec v = 0$ or {\em $\psi$ is a constant of the motion}.

\begin{prop}
If $\psi = \const$., equations of motion (\ref{eq-m+3}) are variational as equations for extremals of the Lagrangian $\Cal L(t, q^l, v^l) {\rm d}t$ on $\R \times \R^3$, such that
\begin{equation}
\Cal L = m_0c^2 \sqrt{\frac{v^2}{c^2} - 1} + \frac{e}{c} \vec{A} \vec{v} - eV. 
\end{equation}
\end{prop}

Finally, let us investigate a Dicke-type influence on tachyons.
Denote $\exp \mu = e^\mu$ and put
\begin{equation}
\mu =  - \frac{\psi}{m_0^2 c^2}, \quad \tilde m_0 = m_0 \, e^\mu.
\end{equation}
An analogous procedure as in the case of ``normal particles" directly leads to the following result:

\begin{prop} \label{3dimT-Dicke}
With help of the mass function $\tilde m_0$, depending upon a scalar potential $\psi$ on $\R^4$, and defined by 
\begin{equation}
 \tilde m_0 = m_0 \exp \left(- \frac{\psi}{m_0^2 c^2}\right),
\end{equation}
equations of motion for tachyons take the following equivalent form:

$\bullet$ in absence of the electromagnetic field:
\begin{equation}
\frac{{\rm d}}{{\rm d}t}\Bigl(\frac{\tilde m_0 \vec{v}}{\sqrt{\frac{v^2}{c^2} - 1}} \Bigr) 
= c^2 \,\sqrt{\frac{v^2}{c^2} - 1}  \, \grad \tilde m_0,
\end{equation}

$\bullet$ in presence of the electromagnetic field:
\begin{equation}
\frac{{\rm d}}{{\rm d}t}\Bigl(\frac{\tilde m_0 \vec{v}}{\sqrt{\frac{v^2}{c^2} - 1}} \Bigr) 
= \frac{\tilde m_0}{m_0} \, \vec{\cal{F}}_L + c^2 \,\sqrt{\frac{v^2}{c^2} - 1}  \, \grad \tilde m_0.
\end{equation}
\end{prop}

We can see that in case of particles with negative square of mass {\em Dicke force} takes the form 
\begin{equation}
{\vec {\Cal F}}_D = c^2 \,\sqrt{\frac{v^2}{c^2} -1}  \, \grad \tilde m_0.
\end{equation}


\section{Particles with zero mass}

It remains to study the case $\Cal M = 0$, i.e. particles with zero mass. Let us again consider 
on $(\R \times T\R^4) \backslash \Cal H_0$ adapted coordinates $(s,q^l,t,v^l,\dot q^4)$, $1 \leq l \leq 3$.

\subsection{The evolution space}

The constraint $\Cal Q \subset (\R \times T\R^4) \backslash \Cal H_0$ is now given by equation
\begin{equation}
(\dot q^4)^2 - \sum_{p=1}^3(\dot{q}^p)^2 = 0, \quad \text{resp.} \quad
(\dot q^4)^2 \left(1-\frac{v^2}{c^2} \right) = 0.
\end{equation}
Since $\dot q^4 \ne 0$, it holds $v = c$, i.e., {\em particles of this kind move with the light speed}. 

Note that in coordinates $(s,q^l,t,v^l,\dot q^4)$, the constraint $\Cal Q_+$ {\em cannot} be expressed
in the form $\dot q^4 = h(s,q^l,t,v^l)$, so that we have to represent the evolution space in the form
\begin{equation}
\Cal Q_+: \quad v=c, \quad \dot q^4 > 0.
\end{equation}
Drawing the constraint $\Cal Q$ and the evolution space $\Cal Q_+$ in the ``velocity space" similarly as in the previous two cases, i.e. supressing the $\dot q^3$ dimension and denoting
$(\dot{q}^1,\dot{q}^2,\dot{q}^4)= c (x,y,w)$, we get for $\Cal Q$
\begin{equation}
w^2 - x^2 - y^2 = 0
\end{equation}
(Fig. 3).
\begin{figure}
\centering
\includegraphics[width=8cm]{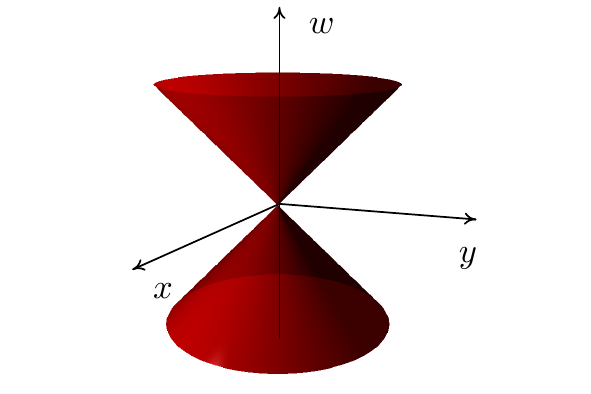}
\caption
{Projection of the evolution space of a massless particle.}
\end{figure}

\subsection{Four-momentum and energy}

Since our definition of  {\em four-momentum} is universal for all particles, we have 
$p = g(\hat u, \cdot)$, and in coordinates 
\begin{equation}
\begin{aligned}
(\hat p_\sigma)_{1 \leq \sigma \leq 4}  &= (- \dot q ^1, - \dot q ^2,- \dot q ^3, \dot q ^4) =
(- \frac{{\vec v} \dot q^4}{c}, \,  \dot q^4),
\\
(p_\sigma)_{1 \leq \sigma \leq 4} &= (- \frac{{\vec v} \dot q^4}{c}, \, c \dot q^4),
\quad \
(p^\sigma)_{1 \leq \sigma \leq 4} = (\frac{{\vec v} \dot q^4}{c}, \, c \dot q^4).
\end{aligned}
\end{equation}

For the {\em equation of the constraint $\Cal Q$}, i.e. $g(\hat u,\hat u) = 0$, we obtain
\begin{equation}
\hat p_\sigma \hat p^\sigma = 0, \quad \text{or} \quad
p_l p^l + \frac{1}{c^2} p_4 p^4 = 0, \quad \text{i.e.} \quad  p_4 p^4 - c^2p^2 = 0,
\end{equation}
where, as usual, $p^2 = \vec p \cdot \vec p = \sum_{l=1}^{3} (p^l)^2$. 

On the evolution space $\Cal Q_+$ it holds $v=c$. Introducing the unit vector $\vec e_v$ in the direction of $\vec v$, i.e.,
\begin{equation}
\vec v = c \, {\vec e}_v,
\end{equation}
we obtain the components of the four-momentum on $\Cal Q_+$ as follows:
\begin{equation}
(p_\sigma)_{1 \leq \sigma \leq 4} = \dot q^4 (- \vec e_v, \, c),
\quad \
(p^\sigma)_{1 \leq \sigma \leq 4} = \dot q^4 (\vec e_v, \, c).
\end{equation}
Let us introduce {\em energy} by the same formula as in the $\Cal M \ne 0$ cases:
\begin{equation}
\Cal E = c\, \dot q^4.
\end{equation}
Then the equation of the constraint $\Cal Q_+$ takes the form $\Cal E^2 - c^2p^2 = 0$, $\Cal E > 0$,
hence, on the evolution space,
\begin{equation}
\Cal E = c \, p.
\end{equation}

\subsection{Equations of motion} 

It is to be stressed that the case $\Cal M = 0$ is much different from the above two, when $\Cal M \ne 0$.
The reason is that the constrained euqations of motion are {\em singular}, in the sense that
they {\em cannot be put into the normal form}. Recall that they read
\begin{equation} 
\begin{aligned}
\Bigr(\delta_{lj} - \frac{\dot{q}^l\dot{q}^j} {\sum_{p=1}^3(\dot{q}^p)^2} \Bigl) \ddot{q}^j 
&= \dot{q}^j\left(\frac{\partial\phi_j}{\partial q^l} -
\frac{\partial\phi_l}{\partial q^j}\right) +
 \frac{\dot{q}^l\dot{q}^j}{\sqrt{\sum_{p=1}^3(\dot{q}^p)^2}} 
\left(\frac{\partial\phi_j}{\partial q^4} -
\frac{\partial\phi_4}{\partial q^j}\right)
\\
& + \sqrt{\sum_{p=1}^3(\dot{q}^p)^2} \, \left(\frac{\partial\phi_4}{\partial
q^l}-\frac{\partial\phi_l}{\partial q^4} \right) -
\frac{\partial\psi}{\partial q^l} - \frac{\partial\psi}{\partial q^4}
\frac{\dot{q}^l}{\sqrt{\sum_{p=1}^3(\dot{q}^p)^2}},
\end{aligned}
\end{equation}
where $1 \leq l \leq 3$, for dynamics proceeding in the evolution space $\Cal Q_{+}$.
\footnote{As it is known, singular equations do not obey the Newton's determinism principle, i.e., by initial conditions in the evolution space a unique solution need not be determined. Lagrangian dynamics of this kind are studied by the Dirac's theory of constrained systems, or by direct methods developed in \cite{Kr-a94} (see also \cite{Kr-a97} and \cite{KruPri}). The latter methods are suitable also for systems that are not Lagrangian, as in our case.}

Let us express these equations in adapted coordinates $(s,q^l,t,v^l,\dot q^4)$. Since along the constrained paths we have
\begin{equation}
\ddot q^j = \frac{{\rm d} \dot q^j}{{\rm d}s} = \frac{1}{c} \frac{{\rm d}( v^j \dot q^4)}{{\rm d}t} 
\frac{{\rm d}t}{{\rm d}s} =  \frac{\dot q^4}{c^2} \left( v^j \frac{{\rm d}\dot q^4}{{\rm d}t} + \dot q^4 \frac{{\rm d}v^j}{{\rm d}t} \right),
\end{equation}
we obtain after some calculations
\begin{equation}
\begin{aligned}
\frac{\dot{q}^4}{c}\left(\delta_{lj}-\frac{v^lv^j}{c^2} \right)\frac{{\rm d}v^j}{{\rm d}t}
&= v^j\left(\frac{\partial\phi_j}{\partial q^l} - \frac{\partial\phi_l}{\partial q^j}\right) +
 \frac{v^l v^j}{c^2} 
\left(\frac{\partial\phi_j}{\partial t} - c \frac{\partial\phi_4}{\partial q^j}\right)
+ \left(c \frac{\partial\phi_4}{\partial q^l} - \frac{\partial\phi_l}{\partial t} \right) 
\\
&- \frac{1}{\dot q^4} \left(
c \frac{\partial\psi}{\partial q^l} + \frac{v^l}{c} \frac{\partial\psi}{\partial t} \right),
\end{aligned}
\end{equation}
where we have used the constraint conditions $v = c$ and $\dot q^4 >0$.

With notations $(\phi_l)_{l=1,2,3} = \frac{e}{c} \vec A$, $ \phi_4 = - \frac{e}{c} V$, and the unit vector 
$\vec e_v$ in the direction of $\vec v$, the motion equations take the vector form as follows:
\begin{equation} \label{prel}
\begin{aligned}
\dot{q}^4 \left[\frac{{\rm d} {\vec e}_v}{{\rm d}t} - \left(\vec e_v \cdot \frac{{\rm d}{\vec e}_v}{{\rm d}t} \right) \, \vec e_v\right]
&= e \, \left(\vec{e}_v \times \rot {\vec A}  - \frac{1}{c}  \frac{\partial \vec A}{\partial t} - \grad V 
 + \Bigl( \vec e_v\cdot \Bigl( \frac{1}{c} \frac{\pr \vec A}{\partial t} + \grad V \Bigr) \Bigr) \, \vec e_v \right)
\\
&- \frac{1}{\dot q^4} \left(
c  \grad \psi + \frac{\partial\psi}{\partial t} \, \vec e_v\right)
\end{aligned}
\end{equation}
(above, the dot means the scalar product of two three-vectors). However, these equations can be simplified. First, we notice that $\vec e_v \cdot \vec e_v = 1$, hence
\begin{equation}
\frac{1}{2} \frac{\rm d}{{\rm d}t} (\vec e_v \cdot \vec e_v) = \vec e_v \cdot \frac{{\rm d} {\vec e}_v}{{\rm d}t} 
 = 0.
\end{equation}
Next, denote
\begin{equation} \label{force}
\vec {\Cal F }=  e \, \left(\vec{e}_v \times \rot {\vec A}  - \frac{1}{c}  \frac{\partial \vec A}{\partial t} - \grad V 
 + \Bigl( \vec e_v\cdot \Bigl( \frac{1}{c} \frac{\pr \vec A}{\partial t} + \grad V \Bigr) \Bigr) \, \vec e_v \right),
\end{equation}
then obviously
\begin{equation}
\vec {\Cal F} \cdot \vec v = \vec {\Cal F} \cdot \vec e_v = 0.
\end{equation}
Now, applying to the motion equation the scalar product with $\vec e_v$, we immediately obtain
\begin{equation}
c \, \vec e_v \cdot \grad \psi  + \frac{\partial\psi}{\partial t}  = \vec v \cdot \grad \psi  + \frac{\partial\psi}{\partial t}  = \frac{{\rm d} \psi}{{\rm d}t} =0.
\end{equation}
(Note that the same result provides eq. (\ref{psider}) where $\Cal M = 0$).

Finally, we notice that the following {\em fields compatibility condition}
\begin{equation}
\vec{\Cal F} \cdot \grad \psi = 0
\end{equation}
has to be satisfied. Indeed, due to the above, equations
\begin{equation}
\dot{q}^4 \frac{{\rm d} {\vec e}_v}{{\rm d}t} = \vec{\Cal F}, \quad \text{and} \quad
c \, \vec e_v \cdot \grad \psi  + \frac{\partial\psi}{\partial t}  = 0
\end{equation}
have to be satisfied simultaneously. Differentiating the latter with respect to $t$ we arrive at
\begin{equation}
\begin{aligned}
0 &= c \,  \frac{{\rm d} {\vec e}_v}{{\rm d}t} \cdot \grad \psi  + c \, \vec e_v \cdot  \frac{\rm d}{{\rm d}t} \grad \psi + 
\frac{\rm d}{{\rm d}t}\frac{\partial\psi}{\partial t}  =
c \,  \frac{{\rm d} {\vec e}_v}{{\rm d}t} \cdot \grad \psi  + c \, \vec e_v \cdot  \grad \frac{{\rm d}\psi}{{\rm d}t}  + 
\frac{\partial}{\partial t} \frac{{\rm d}\psi}{{\rm d}t} 
\\
&= c \,  \frac{{\rm d} {\vec e}_v}{{\rm d}t} \cdot \grad \psi = \frac{c}{\dot q^4} \, \vec{\Cal F} \cdot \grad \psi,
\end{aligned}
\end{equation}
proving our assertion.

Summarising the results, we can conclude:

\begin{prop} \label{3dim-zero}
Let $\phi$ $(\ref{4pot})$ be a one-form (covariant four-vector field), and $\psi$ a function on the Minkowski space-time $\R^4$. Consider on $(\R \times T\R^4) \backslash \Cal H_0$ adapted coordinates 
$(s,q^l,t,v^l, \dot q^4)$ defined by $(\ref{tcoord})$ and $(\ref{vcoord})$.
Extremals of a Lagrangian system defined by the Lagrangian $\Lambda = L {\rm d}s$,
\begin{equation}
L = - \frac{1}{2} \Bigl(1 - \frac{v^2}{c^2} \Bigr) (\dot{q}^4)^2
+ e \, \Bigl(\frac{1}{c} \vec{A} \vec{v} - V \Bigr) \frac{\dot q^4}{c}- \psi,
\end{equation}
and subject to the constraint
\begin{equation} 
v= c, \quad {\dot q}^4 > 0,
\end{equation}
are curves $\c(s) = (t(s),q^l(t(s)))$ in $\R^4$, satisfying the following system of differential equations:
\begin{equation} \label{eq0+3}
\begin{aligned}
\frac{\Cal E}{c} \, \frac{{\rm d} {\vec e}_v}{{\rm d}t} 
&= e \, \left(\vec{e}_v \times \rot {\vec A}  - \frac{1}{c}  \frac{\partial \vec A}{\partial t} - \grad V 
 + \Bigl( \vec e_v\cdot \Bigl( \frac{1}{c} \frac{\pr \vec A}{\partial t} + \grad V \Bigr) \Bigr) \, \vec e_v \right)
\\
\frac{{\rm d} \psi}{{\rm d}t} &\equiv 
c \, \vec e_v \cdot  \grad \psi + \frac{\partial\psi}{\partial t}  = 0,
\end{aligned}
\end{equation}
together with the following equation of the constraint (energy equation):
\begin{equation} \label{Eeq}
c^2\frac{{\rm d}t}{{\rm d}s} = \Cal E > 0.
\end{equation}
As a consequence of these equations, the motion is allowed only if  the vector fields $\vec{\cal F}$ and 
$\grad \psi$ are orthogonal {\em (fields compatibility)}, and proceeds in such a way that the vector fields 
$\vec v$ and ${\rm d} \vec v / \rm{d}t$, $\vec{\Cal F}$ and $\vec v$, and $\grad \psi$ and 
${\rm d} \vec v / \rm{d}t$ are always orthogonal, where 
$\vec v = c \, \vec e_v$, and $\vec{\Cal F}$ is the force on the right-hand side of the first equation.
\end{prop}

Let us study the motion equations in more detail. 

Assume the electric charge $e$ be zero. Then $\vec{\Cal F} = 0$ and the motion is governed by equations
\begin{equation}
\frac{{\rm d} {\vec e}_v}{{\rm d}t}  = 0, \quad 
\frac{{\rm d} \psi}{{\rm d}t} \equiv 
c \, \vec e_v \cdot  \grad \psi + \frac{\partial\psi}{\partial t}  = 0.
\end{equation}
The first equation gives us that $\vec e_v = \vec k = \const.$, i.e. the particle moves in $\R^3$ along a {\em straight line} (with the speed of light). More precisely, every straight line in $\R^3$ is an allowed trajectory, depending upon initial conditions. Thus, it must hold ${\rm d} \psi / {\rm d}t = 0$ along every straight line
in $\R^3$, and consequently, ${\rm d} \psi / {\rm d}t = 0$, i.e. $\psi = const.$ on $\R^3$.

However, this result means that {\em the above two equations are incompatible}. A release is to conclude that {\em massless particles do not feel the scalar field $\psi$}. On the other hand, we have seen earlier (cf. e.g. eq. (\ref{3dimM})) that particles with non-zero mass do feel the field $\psi$. Hence, it turns out that {\em every particle possesses a scalar field charge depending upon the particle's mass, such that  for massless particles the scalar field charge is zero}. With regard to eq. (\ref{3dimM}) 
we may assume this charge being $m_0$, and write
\begin{equation}
\psi = m_0 \bar \psi.
\end{equation}

Now, proposition \ref{3dim-zero} can be reformulated as follows:

\begin{prop} \label{3dim-zero}
\textbf{\em(Equations of motion: particles with zero mass, $3$-dimensional observer.)}
Let $\phi$ $(\ref{4pot})$ be a one-form (covariant four-vector field), and $\psi$ a function on the Minkowski space-time $\R^4$. Consider on $\R \times T\R^4  \backslash \Cal H_0$ adapted coordinates 
$(s,q^l,t,v^l, \dot q^4)$ defined by $(\ref{tcoord})$ and $(\ref{vcoord})$. A massless particle is described by a Lagrangian $\Lambda = L {\rm d}s$,
\begin{equation}
L = - \frac{1}{2} \Bigl(1 - \frac{v^2}{c^2} \Bigr) (\dot{q}^4)^2
+ e \, \Bigl(\frac{1}{c} \vec{A} \vec{v} - V \Bigr) \frac{\dot q^4}{c} - a \psi,
\end{equation}
where the charge $a = 0$, subject to the constraint
\begin{equation} 
v= c, \quad {\dot q}^4 > 0.
\end{equation}
Trajectories are curves $\c(s) = (t(s),q^l(t))$ in $\R^4$, satisfying the following system of differential equations:
\begin{equation} \label{fineq0}
\begin{aligned}
\frac{\Cal E}{c} \, \frac{{\rm d} {\vec e}_v}{{\rm d}t} 
&= e \, \left(\vec{e}_v \times \rot {\vec A}  - \frac{1}{c}  \frac{\partial \vec A}{\partial t} - \grad V 
 + \Bigl( \vec e_v\cdot \Bigl( \frac{1}{c} \frac{\pr \vec A}{\partial t} + \grad V \Bigr) \Bigr) \, \vec e_v \right),
 \\
c^2\frac{{\rm d}t}{{\rm d}s} &= \Cal E > 0,
\end{aligned}
\end{equation}
where $\Cal E$ is a function on $\R^4$, having the meaning of the kinetic energy of the particle.

As a consequence of these equations, the motion proceeds in such a way that the vector fields 
$\vec v$ and ${\rm d} \vec v / \rm{d}t$, and $\vec{\Cal F}$ and $\vec v$ are always orthogonal, where 
$\vec v = c \, \vec e_v$, and $\vec{\Cal F}$ is the force on the right-hand side of the first equation.
\end{prop}

Note that from the form (\ref{fineq0}) of the equations of motion it is explicitly clear that equations for massless particles are indeed singular: they do not provide a uniquely determined motion unless energy of the particle is fixed. In more detail,

\noindent
(i) uncharged massless particles in a scalar field move along straight lines in $\R^3$, with the speed of light, and positive energy.

\noindent
(ii) Charged massless particles in a scalar and electromagnetic field move with the speed of light, the direction of the motion in $\R^3$ is not influenced by the scalar field, is always orthogonal to the electromagnetic force, governed by equation
\begin{equation}
\frac{\Cal E}{c} \, \frac{{\rm d} {\vec e}_v}{{\rm d}t} 
= e \, \left(\vec{e}_v \times \rot {\vec A}  - \frac{1}{c}  \frac{\partial \vec A}{\partial t} - \grad V 
 +\Bigl( \vec e_v\cdot \Bigl( \frac{1}{c} \frac{\pr \vec A}{\partial t} + \grad V \Bigr) \Bigr) \, \vec e_v \right),
\end{equation}
and depends upon a choice of the particle's kinetic energy $\Cal E >0$.

\noindent
In both the cases, {\em a trajectory $\c(s)$ in $\R^4$} can be uniquely determined by initial conditions if energy $\Cal E$ (as a function on $\R^4$) is chosen.

\section*{A brief summary and conclusions}

We proposed a new approach to foundations of relativistic dynamics. It is based on a treatment of a particle as a Lagrangian system subject to a nonholonomic constraint, compatible with the special relativity theory. We explored a geometric setting to nonholonomic mechanics, proposed by one of us \cite{krupkova-JMP}, where the constrained system is considered as ``living" on the submanifold
defined by the constraint: The constraint submanifold thus becomes a genuine {\em evolution space for the constrained system}.

The key idea is that relativistic particles of any mass moving in an electromagnetic field (defined by a four potential $\bar \phi$) and a scalar field (defined by a scalar potential $\bar \psi$) can be described by 

\smallskip

\noindent
(1) the Lagrange function
\begin{equation}
L = - \frac{1}{2} g_{\sigma \nu} \dot q^\sigma \dot q^\nu + e \bar \phi_\sigma \dot q^\sigma - a \bar \psi,
\end{equation}
defined on $\R \times T\R^4$, where $g$ is the Minkowski metric on $\R^4$, $g = (-1,-1,-1,1)$, and
$e$, resp.  $a$, is the particle's charge corresponding to the electromagnetic and scalar field, respectively,

\smallskip

\noindent
and

\smallskip

\noindent
(2) the nonholonomic constraint 
\begin{equation}
g_{\sigma \nu} \dot q^\sigma \dot q^\nu  = \Cal M c^2
 \end{equation}
in $\R \times T\R^4$, where $\Cal M$ is a function on the space-time, representing the particle's mass.

\smallskip

As a main result, {\em we found equations of motion as they appear to a four-dimensional, and to a three-dimensional observer}.

\smallskip

We discussed in detail the cases when $\Cal M = \const. = m_0^2$ (usual particles), $\Cal M = \const. = - m_0^2$ (tachyons), and $\Cal M = 0$ (massless particles). 

As expected, {\em massless} particles move with the light speed, and particles with a {\em positive square of mass} move with a speed lower than the speed of light;  the constant $m_0$ has the meaning of rest mass. On the other hand, particles with a {\em negative square of mass} (tachyons) are particles with {\em real} positive mass $m_0$, moving with a speed {\em greater than the light speed}. The constant $m_0$ is then the mass of a tachyon moving with the speed $c\sqrt{2}$. (Alternatively, tachyons can be understood as particles with imaginary rest mass $im_0$ moving with a speed lower than the speed of light, if modeled on a {\em complex} Minkowski space).

It turned out that the scalar field charge $a$ is closely connected with the mass of the particle: for massless particles $a = 0$, and $a \ne 0$ in the other cases. This means that {\em the presence of a scalar field does not effect the motion of massless particles}. On the other hand, all massive particles are influenced by a scalar field. We have found that the corresponding force takes the form
for ``normal" particles, respectively for tachyons: 
\begin{equation}
\vec{\Cal F}_D = - c^2 \sqrt{1- \frac{v^2}{c^2}} \, \grad \tilde m_0, \quad \text{respectively} \quad
\vec{\Cal F}_D = c^2 \sqrt{\frac{v^2}{c^2} - 1} \, \grad \tilde m_0,
\end{equation}
where
\begin{equation}
\tilde m_0 = m_0 \exp \left(\frac{\bar \psi}{m_0c^2} \right), \quad \text{respectively,} \quad
\tilde m_0 = m_0 \exp \left(- \frac{\bar \psi}{m_0c^2} \right),
\end{equation}
and we have put $a = m_0$. 

We called $\vec{\Cal F}_D$ {\em Dicke force} in honour of Dicke who predicted existence of a force of this kind \cite{dicke}. 

The charge $e$ has the usual meaning, however, the corresponding force acting on massive and massless particles is different. For massive particles we get the usual formula for the Lorentz force (the same for ``normal" particles and tachyons).
For massless particles the corresponding force takes the form
\begin{equation}
\vec{\Cal F} = e \, \left(\vec{e}_v \times \rot {\vec A}  - \frac{1}{c}  \frac{\partial \vec A}{\partial t} - \grad V 
 + \vec e_v\cdot \Bigl( \frac{1}{c} \frac{\pr \vec A}{\partial t} + \grad V \Bigr) \, \vec e_v \right).
\end{equation}

It is important to stress a striking difference between dynamics of massive and massless particles. Massive particles are {\em regular} mechanical systems in the sense that the dynamics in the evolution space obey the Newton's determinism principle (motion equations can be put into the normal form, i.e. every trajectory is uniquely determined by initial conditions). On the other hand, massless particles are {\em singular} mechanical systems: the dynamics in the evolution space do not obey the Newton's determinism principle (motion equations cannot be put into the normal form, i.e. trajectories are not completely determined by initial conditions). To get a concrete trajectory in the space-time $\R^4$, one
has to fix initial conditions and the {\em energy} of the particle.

We also introduced the four-momentum of a particle as the one form, canonically related with the $2$-tensor field $g$. With help of four momentum $\hat p = (\hat p_\sigma)$, the evolution space (= the constraint submanifold)
is given by equation
\begin{equation}
\hat p_\sigma \hat p^\sigma = m_0^2 c^2, \quad \text{respectively,} \quad
\hat p_\sigma \hat p^\sigma = - m_0^2 c^2, \quad \text{respectively,} \quad
\hat p_\sigma \hat p^\sigma = 0
\end{equation}
for ``normal" particles, tachyons, and massless particles, respectively. 
On the evolution space components of the four-momentum were shown to be as follows:
\begin{equation}
(\hat p_\sigma) = (-m\vec v, \, \Cal E/c), \quad \text{respectively,} \quad
(\hat p_\sigma) = (- \frac{\Cal E \vec v}{c^2}, \, \Cal E/c) = \frac{\Cal E}{c} (- \vec e_v, \ 1)
\end{equation}
for massive, respectively, massless particles, where 
\begin{equation}
m = \frac{m_0c}{\sqrt{1 - \frac{v^2}{c^2}}}, \quad \text{respectively,} \quad
m = \frac{m_0c}{\sqrt{\frac{v^2}{c^2} - 1}},
\end{equation}
for ``normal" particles, respectively, for tachyons, and $\Cal E$ is the kinetic energy of the particle,
defined (in all the cases) by formula
\begin{equation}
\Cal E = c\dot q^4.
\end{equation}
Consequently, on the evolution space, energy is given by the following formulas:
\begin{equation}
\Cal E = c \sqrt{p^2 + m_0^2c^2}, \quad \text{respectively,} \quad
\Cal E = c \sqrt{p^2 - m_0^2c^2}, \quad \text{respectively,} \quad \Cal E = cp,
\end{equation}
for ``normal" particles, tachyons, and massless particles, respectively,
where $p = \sqrt{\vec p \cdot \vec p}$ is the usual length of the momentum three-vector $\vec p$.

Finally let us mention variationality aspects of the obtained equations of motion.
In the particular case of ``usual" particles with positive square of mass moving in an {\em electromagnetic field}, the ``$(3+1)$-dimensional" equations of motion we have obtained, take the usual form. Hence, as it is known, they can be obtained from a variational principle as Euler--Lagrange equations of the Lagrangian (\ref{standL}). On the other hand, rather surprisingly, in presence of a scalar field, the obtained equations {\em do not come from} an (unconstrained) {\em variational principle}.\footnote{The reader might be interested that, however, equations of motion we have obtained in this paper are ``variational" in a generalized sense: one can find a {\em nonholonomic variational principle} providing these equations (see \cite{Kru-09}).} In particular, it might be interesting that trying a ``natural" extension of the Lagrangian (\ref{standL}) by adding to the Lagrangian a scalar field term, does not provide the correct equations of motion (recall Sec. \ref{secvar}).

\section*{Acknowledgements}

Research supported by grants GACR 201/06/0922 and 201/09/0981
of the Czech Science Foundation, and MSM 6198959214 and MSM 0021622409
of the Ministry of Education, Youth and Sports of the Czech Republic.


\begin{thebibliography}{99}

\bibitem{bransdicke} C. Brans and R.H. Dicke,
Mach's Principle and a Relativistic Theory of Gravitation,
 {\em Phys. Rev.} \textbf{124} (1961) 925--935.

\bibitem{CaRa}  J.F. Cari\~ nena and M.F. Ra\~ nada,
Lagrangian systems with constraints: a geometric approach to the method of 
Lagrange multipliers,
{\em J. Phys. A: Math. Gen.} \textbf{26} (1993) 1335--1351.

\bibitem{Chet} 
N.G. Chetaev,  
On the Gauss principle,
{\em Izv. Kazan. Fiz.-Mat. Obsc.} \textbf {6} (1932-33) 323--326 (in Russian).

\bibitem{leon-et-al} 
M. de Le\'{o}n, J. C. Marrero, and D. M. de Diego, Non-holonomic Lagrangian systems in jet manifolds, 
{\em J. Phys. A: Math. Gen.} \textbf{30} (1997) 1167--1190.

\bibitem{dicke} 
R. Dicke, The influence of the time dependent gravitation
interaction on the Solar system. In: {\em Gravity and Relativity}, Ed. W. F. Hoffmann, Mir, Moscow, 1965
(in Russian).

\bibitem{Gi}  G. Giachetta, 
Jet methods in nonholonomic mechanics,
{\em J. Math. Phys.} \textbf{33} (1992) 1652--1665.

\bibitem{Hel}
H. Helmholtz, 
Ueber die physikalische Bedeutung des Prinzips der kleinsten Wirkung, 
{\it J. f\"ur die reine u. angewandte Math.} \textbf{100} (1887) 137--166.

\bibitem{KoMa}  W.S. Koon and J.E. Marsden, 
The Hamiltonian and Lagrangian approaches to the dynamics of
nonholonomic systems,
{\em Rep. Math. Phys.} \textbf{40}  (1997) 21--62.

\bibitem{Kr73} 
D. Krupka,
Some geometric aspects of variational problems in fibered manifolds,
{\em Folia Fac. Sci. Nat. Univ. Purk. Brunensis}, Physica \textbf{14}, Brno, 
Czechoslovakia, 1973, 65pp.; ArXiv:math-ph/0110005.

\bibitem{Kr08} 
D. Krupka,
Global variational theory in fibred spaces, 
in: {\it Handbook of Global Analysis}, Elsevier, 2008, 755--839.

\bibitem{Kr-a94}
O. Krupkov\'a, 
A geometric setting for higher-order Dirac-Bergmann theory of constraints, 
{\em J. Math. Phys.} \textbf{35} (1994) 6557-6576.

\bibitem{Kr-a97} 
O. Krupkov\'a, 
{\em The Geometry of Ordinary Variational Equations}, 
Lecture Notes in Mathematics \textbf{1678}, Springer, Berlin, 1997.

\bibitem{krupkova-JMP} 
O. Krupkov\'{a}, Mechanical systems with non-holonomic
constraints, {\em J. Math. Phys.} \textbf{38} (1997)
5098--5126.

\bibitem{krupkova-RMP}
O. Krupkov\'{a}, Recent results in the geometry of constrained systems. {\em Rep. Math. Phys.} 
\textbf{49} (2002) 269--278.

\bibitem{Kru-09}
O. Krupkov\'{a}, The nonholonomic variational principle. {\em J. Phys. A: Math. Theor.}, in print.

\bibitem{krupkova-musilova-rel}
O. Krupkov\'{a} and J. Musilov\'{a}, The relativistic particle as a
mechanical system with non-holomic constraints, {\em J. Phys. A: Math. Gen.} \textbf{34} (2001) 3859--3875.

\bibitem{KruPri}
O. Krupkov\'a and G.E. Prince, 
Second Order Ordinary Differential Equations in Jet Bundles and the Inverse Problem of the Calculus of Variations, 
In: {\em Handbook of Global Analysis} (Elsevier, 2008) 841-908.

\bibitem{massa-pagani}
E. Massa and E. Pagani, Classical mechanics of non-holonomic systems: a geometric approach. {\em Ann. Inst. Henri Poincar\'{e}} \textbf{66} (1997) 1--36.

\bibitem{Nov} 
J. Novotn\'y, On the inverse variational problem in the classical mechanics,
In: {\em Proc. Conf. on Diff. Geom. and Its Appl. 1980}, O. Kowalski, ed.
(Universita Karlova, Prague, 1981) 189--195.

\bibitem{sarlet} 
W. Sarlet, 
A direct geometrical construction of the dynamics of non-holonomic Lagrangian systems, 
{\em Extracta Mathematicae} \textbf{11} (1996) 202--212.

\bibitem{sarlet-et-al-1}
W. Sarlet, F. Cantrijn and D.J. Saunders, A geometrical framework for the study of non-holonomic Lagrangian systems, {\em J. Phys. A: Math. Gen.} \text{28} (1995) 3252--3268.

\bibitem{sarlet-et-al-2}
D.J. Saunders, W. Sarlet and F. Cantrijn, A geometrical framework for the study of non-holonomic Lagrangian systems II, {\em J. Phys. A: Math. Gen.} \textbf{29} (1996) 4265--4274.


\end{thebibliography}
\end{document}